\documentclass{aa}  

\usepackage[multidot]{grffile}
\usepackage{tikz}
\usepackage{subfig}
\renewcommand{\thesubfigure}{\Alph{subfigure}}
\usepackage[labelfont=bf]{caption}
\usepackage{natbib}
\usepackage{graphicx}
\usepackage{txfonts}
\usepackage[normalem]{ulem}

\usepackage{csquotes}
\usepackage[makeroom]{cancel}

\usepackage{scalerel,stackengine}
\stackMath
\newcommand\reallywidehat[1]{%
\savestack{\tmpbox}{\stretchto{%
  \scaleto{%
    \scalerel*[\widthof{\ensuremath{#1}}]{\kern-.6pt\bigwedge\kern-.6pt}%
    {\rule[-\textheight/2]{1ex}{\textheight}}
  }{\textheight}%
}{0.5ex}}%
\stackon[1pt]{#1}{\tmpbox}%
}

\begin{document}

   \title{Spectral properties and energy transfer at kinetic scales in collisionless plasma turbulence}

   \subtitle{}

   \author{G. Arrò
          \inst{1,2}
          \and
          F. Califano
          \inst{2}
          \and
          G. Lapenta
          \inst{1}
          }

   \institute{Department of Mathematics, KU Leuven, Leuven, Belgium\\
             \email{giuseppe.arro@kuleuven.be}
             \and
             Dipartimento di Fisica "E. Fermi", Università di Pisa, Pisa, Italy
             }

   \date{}

 
  \abstract
   {Recent satellite observations in the solar wind and in the Earth's magnetosheath have shown that the turbulent magnetic field spectrum, that is know to steepen around ion scales, has another break at electron scales, where it becomes even steeper. The origin of this second spectral break is not yet fully understood and the shape of the magnetic field spectrum below electron scales is still under debate.
   }
   {By means of a fully kinetic simulation of freely decaying plasma turbulence, we study the spectral properties and the energy exchanges characterizing the turbulent cascade in the kinetic range.
   }
   {We start by analyzing the magnetic field, electron velocity and ion velocity spectra at fully developed turbulence. We then investigate the dynamics responsible for the development of the kinetic scale cascade by analyzing the ion and electron filtered energy conversion channels, represented by the electromagnetic work ${\textbf{J}\!\cdot\!\textbf{E}}$, pressure-strain interaction ${-\textbf{P}\!:\!\nabla\,\textbf{u}}$ and the cross-scale fluxes of electromagnetic (e.m.) energy and fluid flow energy, accounting for the nonlinear scale-to-scale transfer of energy from large to small scales.
   }
   {We find that the magnetic field spectrum follows the  ${k^{-\alpha}\,exp(-\lambda\, k)}$ law at kinetic scales with $\alpha\!\simeq\!2.73$ and $\lambda\!\simeq\!\rho_e$ (where ${\rho_e}$ is the electron gyroradius). The same law with $\alpha\!\simeq\!0.94$ and $\lambda\!\simeq\!0.87\rho_e$ is observed in the electron velocity spectrum but not in the ion velocity spectrum that drops like a steep power law $\sim k^{-3.25}$ before reaching electron scales. By analyzing the filtered energy conversion channels, we find that the electrons play a major role with respect to the ions in driving the magnetic field dynamics at kinetic scales. Our analysis reveals the presence of an indirect electron-driven mechanism that channels the e.m. energy from large to sub-ion scale more efficiently than the direct nonlinear scale-to-scale transfer of e.m. energy. This mechanism consists of three steps: in the first step the e.m. energy is converted into electron fluid flow energy at large scales; in the second step the electron fluid flow energy is nonlinearly transferred towards sub-ion scales; in the final step the electron fluid flow energy is converted back into e.m. energy at sub-ion scales. This electron-driven transfer drives the magnetic field cascade up to fully developed turbulence, after which dissipation becomes dominant and the electrons start to subtract energy from the magnetic field and dissipate it via the pressure-strain interaction at sub-ion scales. 
   }
   {}


   \maketitle
%

\section{Introduction}

Turbulence in collisionless magnetized plasmas is a process involving the nonlinear transfer of energy across a wide range of scales, extending from large fluid magnetohydrodynamic (MHD) scales, where the energy is typically injected, down to ion and electron kinetic scales, associated with different physical regimes. Such a complex turbulent cascade still lacks a complete theoretical description and plasma turbulence is mainly studied by means of numerical simulations and satellite measurements conducted in the solar wind (SW) and in the Earth's magnetosphere \citep{BrunoCarbone, Matthaeus2021}. 

The multiscale nature of plasma turbulence is reflected in the shape of the turbulent spectra that exhibit different behavior in different ranges of scales. SW \textit{in situ} observations show that the magnetic field spectrum follows a power law at large MHD scales, with a scaling exponent ranging between $-3/2$ and $-5/3$ depending on certain conditions such as the SW speed and the heliocentric distance \citep{biskamp_2003,chen2013residual,chen2020evolution}. Similar MHD scale power laws are also observed in the magnetic spectra measured in the Earth's magnetosheath, although they are usually shallower than $k^{-3/2}$ in regions close to the bow shock and tend to the Kolmogorov-like $\sim\! k^{-5/3}$ spectrum when moving towards the flanks of the magnetopause \citep{huang2017existence,Stawarz_2019}. Both SW and magnetosheath measurements show that the magnetic field spectrum breaks and steepens at ion scales. Here a different power law develops, with a scaling exponent that varies between $-2$ and $-4$ close to the transition from MHD to sub-ion scales and seems to tend to $\sim\! k^{-2.8}$ when approaching electron scales \citep{Alexandrova_2008,Alexandrova2013,Bourouaine2012,bruno2014spectral,stawarz2016observations,li2020evolution}. Electron scale satellite measurements are harder to obtain but relatively recent observations reveal the presence of a second break and steepening in the magnetic spectra at scales of the order of the electron gyroradius $\rho_e$ \citep[e.g.][]{Alexandrova2009,Alexandrova2012,Alexandrova2013,Alexandrova2021, Sahraoui2009, Sahraoui2010,sahraoui2013scaling,huang2014kinetic,chen2017nature,macek2018magnetospheric}. The shape of the magnetic spectrum at electron scales is still under debate and different descriptions have been proposed and tested on satellite data under different conditions. \citet{Alexandrova2012} have shown that at scale smaller than ion kinetic scales, down to $\rho_e$, the magnetic energy spectra measured in the SW can be described by a law of the form $\sim\! k^{-\alpha}\,exp(-\lambda\, k)$, with a scaling exponent $\alpha\!\sim\!2.8$ and a characteristic length $\lambda$ of the order of $\rho_e$. This scaling, dubbed the \textit{exp} model, has been tested on a large number of magnetic spectra measured at various distances from the Sun, ranging from $0.3$ to $1$ AU \citep{Alexandrova2021}. The presence of an exponential decay in the magnetic spectrum, which reminds of the exponentially decreasing dissipation range of hydrodynamic turbulence \citep{Chen1993}, has been suggested to indicate the onset of dissipation that may actually take place at electron scales in collisionless plasma turbulence, with $\rho_e$ playing the role of the dissipation scale \citep{Alexandrova2013, Alexandrova2012, Alexandrova2021}. On the other hand, many cases have been reported in which the shape of the magnetic spectrum below electron scales is better represented by a power law. In \citet{sahraoui2013scaling} and \citet{huang2014kinetic} a double power law model is used to fit a large number of magnetic spectra at kinetic scales in the SW and  in the Earth's magnetosheath respectively. A scaling consistent with $\sim\! k^{-2.8}$ is found above the electron scale break while the scaling exponent shows a broad variation at electron scales, with steeper slopes in the case of the magnetosheath with respect to the SW. Because of such variations the authors suggest that the scaling of turbulence may not be universal at electron scales, even though the scale of the spectral break shows a strong correlation with $\rho_e$. A power law scaling at electron scales is also found in \citet{chen2017nature} where the authors compare magnetosheath data with a theoretical model based on inertial kinetic Alfvén waves, finding a good agreement between the measured scaling exponent and the one predicted by the model. However, the shape of the magnetic field spectrum at electron scales has been shown to be influenced by many factors related to instrumental limitations \citep{sahraoui2013scaling,huang2014kinetic} and by the presence of whistler waves superimposed to the underlying turbulence \citep{matteini2017electric,lacombe2014whistler,roberts2017variability}. Therefore, as of today there is no general agreement about the shape of the magnetic spectrum at electron scales.

Spectral breaks have been observed and studied also in numerical simulations of plasma turbulence. The transition from MHD to sub-ion scales has been examined using both hybrid \citep[e.g.][]{franci2015high,Franci2016,Franci2017,franci2020modeling,cerri2016subproton,cerri2017reconnection,cerri2017kinetic,cerri2018dual,Servidio2015} and fully kinetic \citep[e.g.][]{roytershteyn2015generation,parashar2018dependence,grovselj2018fully,gonzalez2019turbulent,cerri2019kinetic,pecora2019statistical,rueda2021three,Adhikari_2021} simulations. A general agreement with satellite measurements is found at ion scales, with magnetic spectra whose shape is overall consistent with the observed ion scale power laws. On the other hand, due to computational limitations, electron scales are harder to investigate also in simulations. Nevertheless, a steepening in the magnetic spectrum at electron scales has been observed and discussed in fully kinetic simulations of turbulence under different conditions. In \citet{camporeale2011dissipation,chang2011whistler,gary2012forward,rueda2021three,Franci_2022} the authors observe a transition from ion to electron scales consistent with a double power law model while in \citet{roytershteyn2015generation} the observed magnetic spectrum is well approximated by the \textit{exp} model. Finally, in \citet{parashar2018dependence} the effects of plasma $\beta$ (ratio between the total kinetic pressure and the magnetic pressure) on the turbulence are studied and the authors find that the magnetic spectrum steepens at electron scales, with a curvature that increases with increasing $\beta$.

Many attempts have been made to model and explain the spectral steepening at ion and electron kinetic scales, often using reduced models based on specific processes such as wave-particle interactions \citep{Schekochihin_2009,Howes2011, Boldyrev_2013, TenBarge_2013, Schreiner_2017}, instabilities and magnetic reconnection \citep{Franci2017, Loureiro_2017}. A complementary approach to study the turbulent cascade consists in analyzing the channels responsible for the exchanges between different forms of energy and their relative importance at different scales \citep{Yang2017_2, Yang2017_1, Pezzi2021}. This method, largely employed in hydrodynamic turbulence, was recently applied to plasma turbulence and follows directly from the analysis of the equations for the low-pass filtered fluid flow energy $E^<_{f,s}$ (with $s$ indicating the species) and electromagnetic (e.m.) energy $E^<_{e.m.}$ \citep{Matthaeus_2020}:

\begin{gather}
\partial_t E^<_{f,s}+\nabla \cdot \textbf{J}^u_s=-\Pi_s^{uu}-PS^<_s+W^<_s \label{T1}
\\ \notag \\
\partial_t E^<_{e.m.}+\nabla\cdot \textbf{J}^b=-\sum_s\Pi^{bb}_s-\sum_s W^<_s \label{T2}
\end{gather}
\\with $E^<_{f,s}\!=\!m_s\,\overline{n}_s\,|\widehat{\textbf{u}}_s|^2/2$ and $E^<_{e.m.}\!=\!(|\overline{\textbf{E}}|^2+|\overline{\textbf{B}}|^2)/(8\pi)$, where $m_s$, $n_s$ and $\textbf{u}_s$ are the mass, number density and fluid velocity of particles of species $s$, $\textbf{E}$ and \textbf{B} are the electric and magnetic fields respectively, the bar $\,\overline{\cdot}\,$ indicates the low-pass filtering operation, while the hat $\,\widehat{\cdot}\,$ is the density-weighted filtering (i.e. $\widehat{q}\!=\!\overline{nq}/\overline{n}$ for a generic quantity $q$, with $n$ being the density) \citep{favre1969statistical}. $\textbf{J}^u_s$ and $\textbf{J}^b$ are the low-pass filtered fluid flow energy and e.m. energy fluxes respectively, accounting for the spatial transport of energy. $\Pi_s^{uu}$ and $\Pi^{bb}_s$ represent the energy flux from large to small scales referred to the fluid flow energy and to the e.m. energy respectively. The terms actually describing the exchanges between different forms of energy are the low-pass filtered e.m. work done on the particles $W^<_s\!=\!\overline{\textbf{J}}_s\cdot\widehat{\textbf{E}}$ (where $\textbf{J}_s\!=\!q_s\,n_s\,\textbf{u}_s$ is the electric current density of species $s$, with charge $q_s$) and the low-pass filtered pressure-strain interaction $PS^<_s\!=\!-\overline{\textbf{P}}_s\!:\!\nabla\,\widehat{\textbf{u}}_s\!=\!-\overline{P}_{s,nm}\partial_m \widehat{u}_{s,n}$ (where $\textbf{P}_s$ is the pressure tensor of species $s$ and $\nabla\,\textbf{u}_s\!=\!\partial_m u_{s,n}$ is the strain tensor, containing the derivatives of $\textbf{u}_s$), the latter accounting for the conversion of fluid flow energy into internal (thermal) energy \citep{DelSarto2017}. This scale filtering technique applied to numerical simulations reveals that when the turbulence is fully developed, $W^<_s$ is mainly dominant at scales of the order of a few ion inertial lengths $d_i$. On the other hand, $PS^<_s$ becomes important at sub-ion scales, thus showing that the e.m. energy is transferred to the fluid flow energy at relatively large scales and finally converted into internal energy at kinetic scales, with $\Pi_s^{uu}$ acting as a bridge between these two conversions taking place at different scales \citep{Yang2017_2, Yang2018,  Matthaeus_2020}.

In this work we study the spectral properties of the turbulent cascade at kinetic scales by means of a fully kinetic particle-in-cell (PIC) simulation of freely decaying plasma turbulence and we investigate the development of such spectral features in terms of the filtered energy conversion channels previously described.

\section{Simulations setup}

The simulation was realized using the energy conserving semi-implicit PIC code ECsim \citep{LAPENTA2017, Markidis2011}. We consider a 2D square periodic domain of size $L\!=\!64\,d_i$, with $2048^2$ grid points and $5000$ particles per cell for each of the two species considered, ions and electrons. The ion-to-electron mass ratio is $m_i/m_e\!=\!100$, corresponding to an electron inertial length $d_e \! = \! 0.1 \,d_i$. Both species are initialized using a Maxwellian distribution function with uniform density, uniform and isotropic temperature ($T_\perp\!=\!T_\parallel$) and plasma beta equal to $\beta_i\!=\!8$ for the ions and $\beta_e\!=\!2$ for the electrons (such parameters are chosen in order to reproduce conditions similar to those met in the Earth's magnetosheath, see for example \citet{Phan2018, Stawarz_2019, Bandyopadhyay2020}). With these values for $\beta_i$ and $\beta_e$, the ion and electron gyroradii are equal to $\rho_i\!=\!\sqrt{\beta_i}\,d_i\!\simeq\!2.83\,d_i$ and $\rho_e\!=\!\sqrt{\beta_e}\,d_e\!\simeq\!1.41\,d_e$ respectively at the beginning of the simulation. An uniform out-of-plane guide field $\textbf{B}_0$ is present and the turbulence is triggered by random phase isotropic magnetic field and velocity perturbations with wavenumber $k$ in the range $1\! \leqslant \!k/k_0\! \leqslant \!4$ (with $k_0\!=\!2\pi/L$). In particular, we consider magnetic field fluctuations $\delta \textbf{B}$ with root mean square amplitude $\delta B_{rms}/B_0\!\simeq\!0.9$ and velocity fluctuations $\delta \textbf{u}$ with $\delta u_{rms}/c_A\!\simeq\!3.6$ (where $c_A$ is the Alfvén speed) for both the ions and the electrons. The injection of energy close to ion scales is also justified by the fact that we want to reproduce conditions similar to those of the Earth's magnetosheath, where the correlation length of turbulent fluctuations is typically much smaller than in the SW. The ratio between the plasma frequency and the cyclotron frequency is $\omega_{p,i}/\Omega_i\!=\!100$ for the ions and $\omega_{p,e}/\Omega_e\!=\!10$ for the electrons. The time step used to advance the simulation is $\Delta t\!=\!0.05\, \Omega^{-1}_e$ (where $\Omega_e$ is the electron cyclotron frequency).

\section{Results}

\begin{figure}[t]
\centering
\subfloat[]{
\includegraphics[width=.9\linewidth]{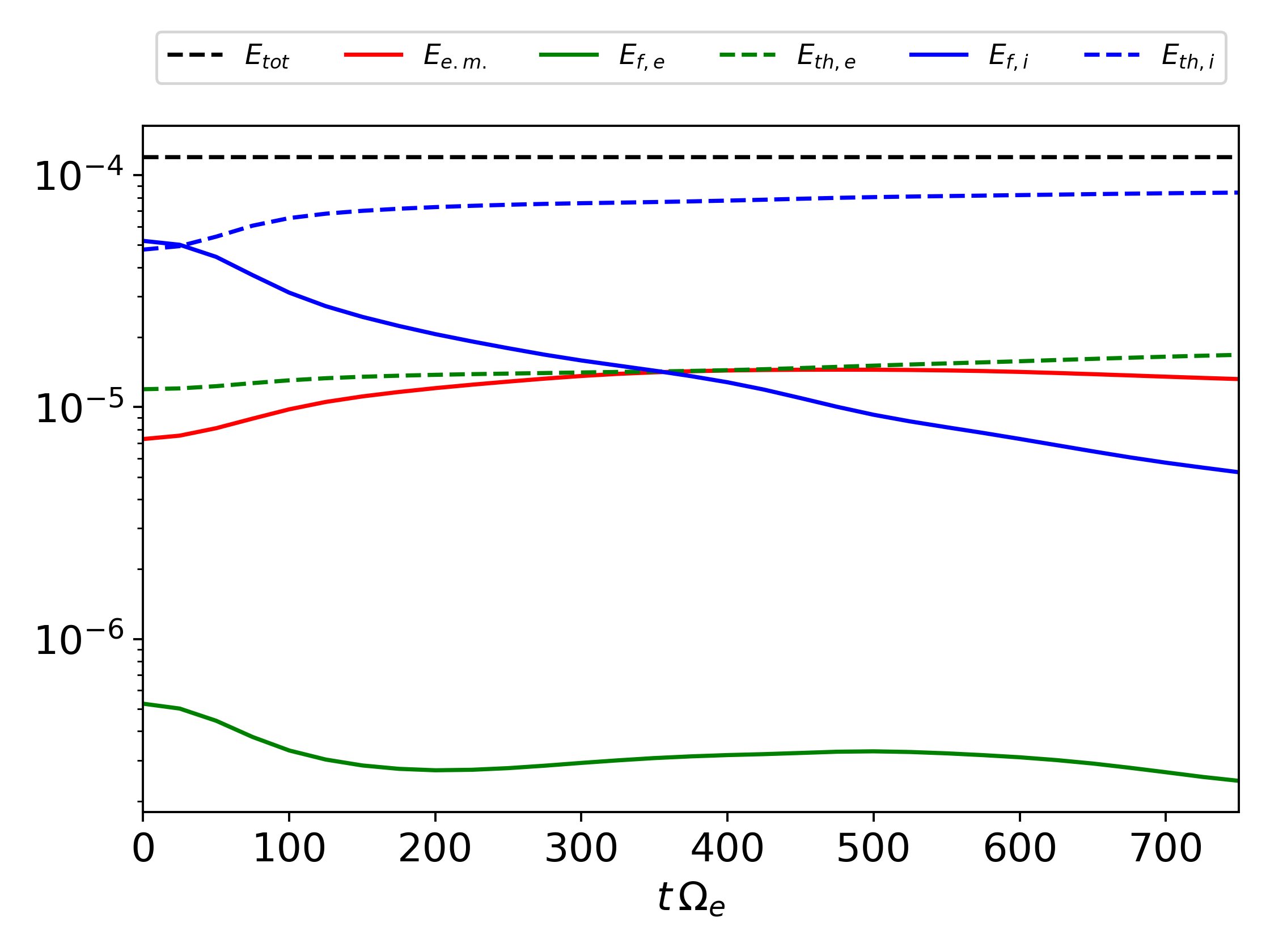} 
\label{Energy}
}
\\
\subfloat[]{
\includegraphics[width=.9\linewidth]{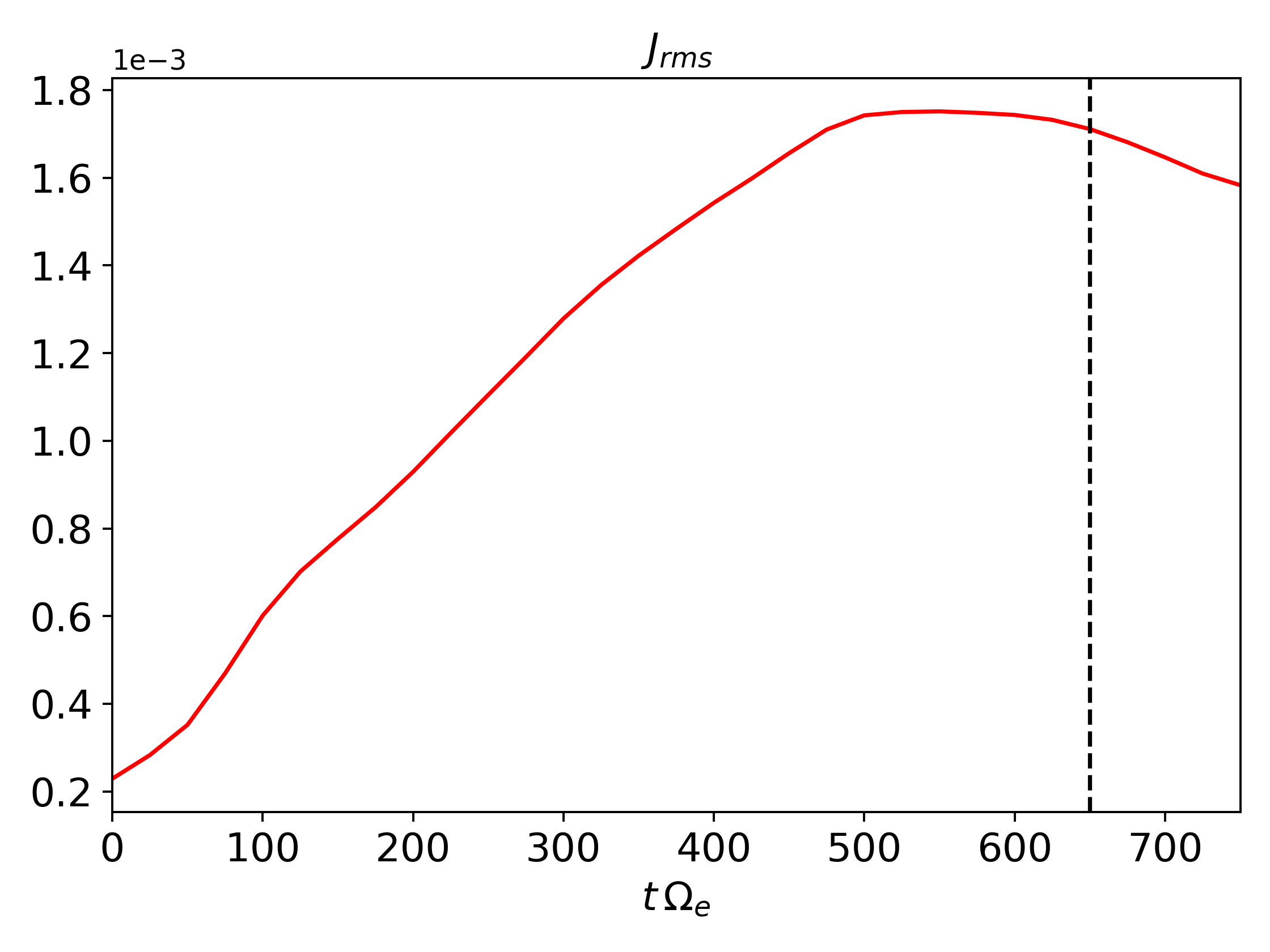} 
\label{Jrms}
}
\caption{(A) Time evolution of total energy $E_{tot}$, magnetic energy $E_{e.m.}$, electron and ion fluid flow energies $E_{f,e}$ and $E_{f,i}$, electron and ion thermal energies $E_{th,e}$ and $E_{th,i}$. (B) Root mean square of the total current $J_{rms}$ as a function of time, where the vertical dashed line indicates the time $t\!=\!650\,\Omega^{-1}_e$.}
\label{F-1}
\end{figure}

\begin{figure*}[t]
\centering
\subfloat[]{
\includegraphics[width=.48\linewidth]{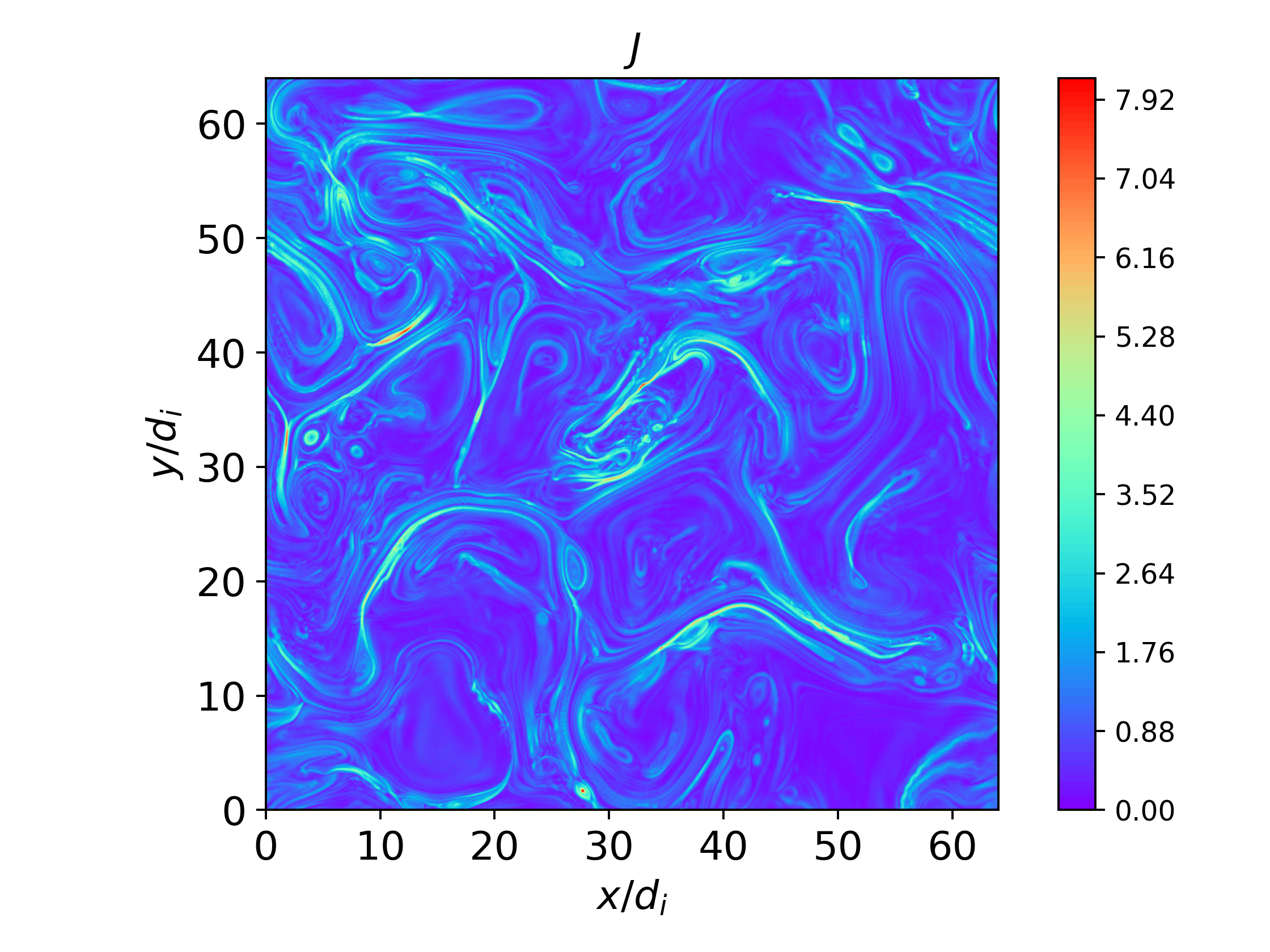} 
\label{J}
}
\hfill
\subfloat[]{
\includegraphics[width=.48\linewidth]{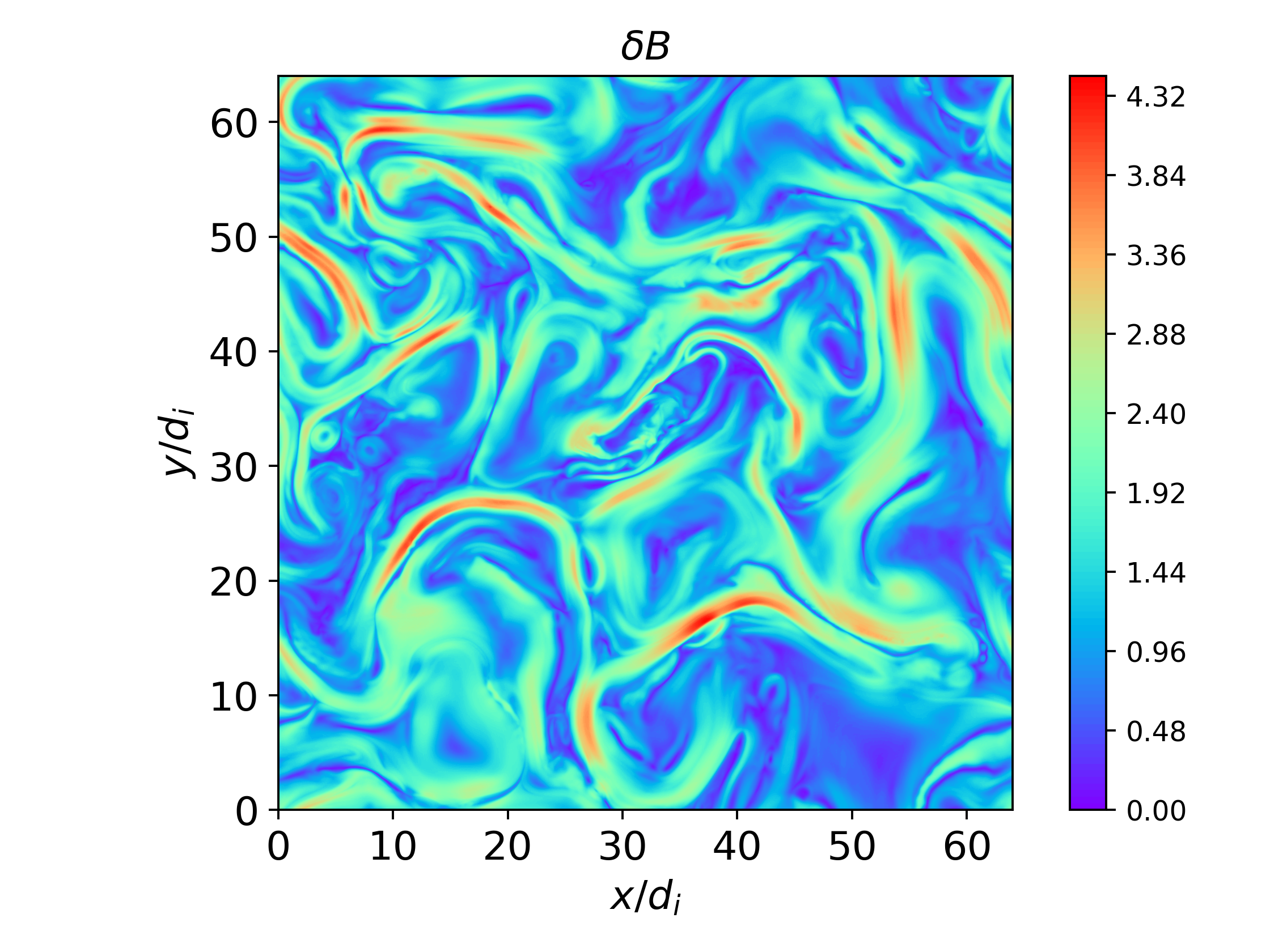}
\label{B}
}
\\
\subfloat[]{
\includegraphics[width=.48\linewidth]{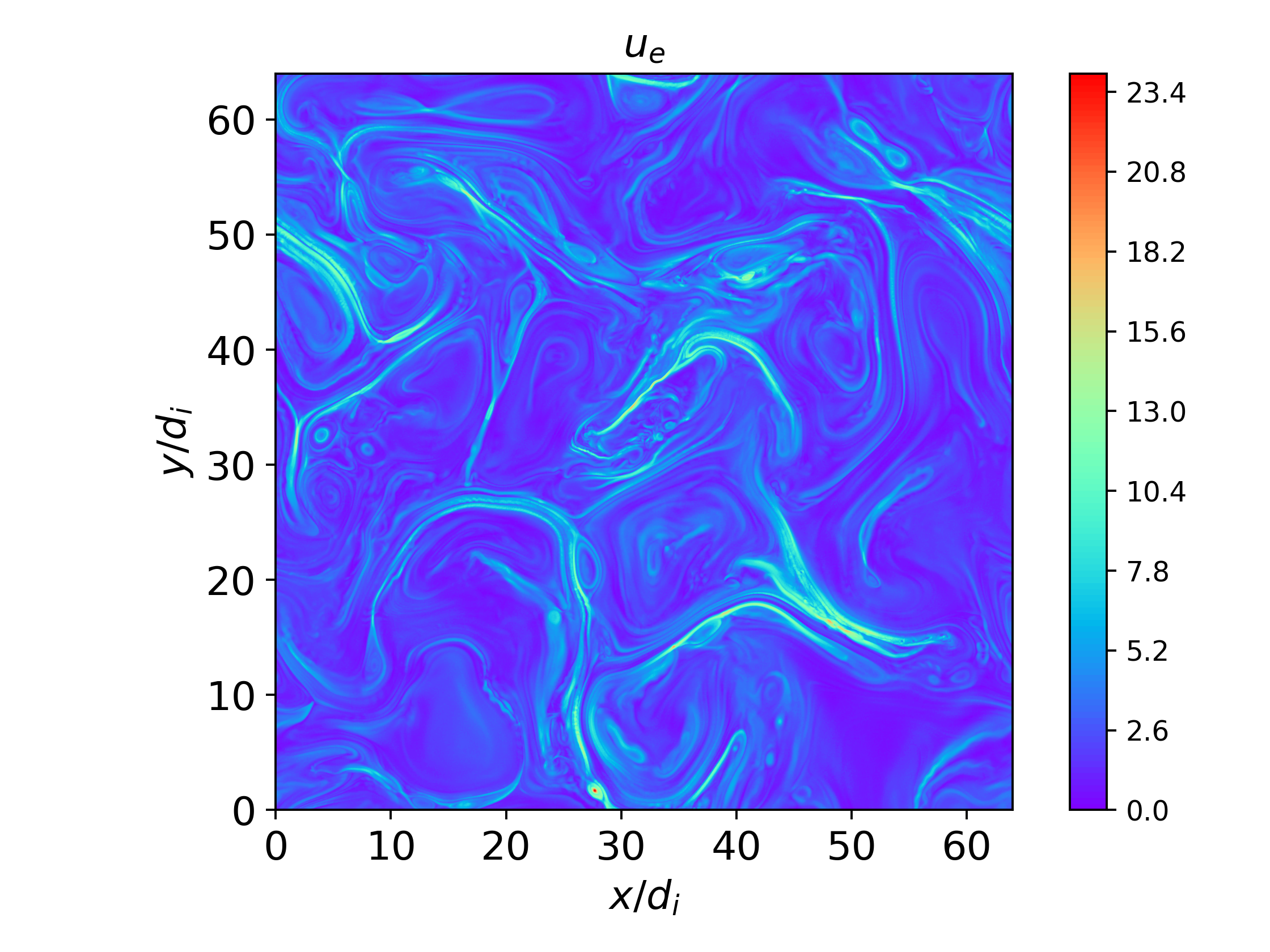} 
\label{u_e}
}
\hfill
\subfloat[]{
\includegraphics[width=.48\linewidth]{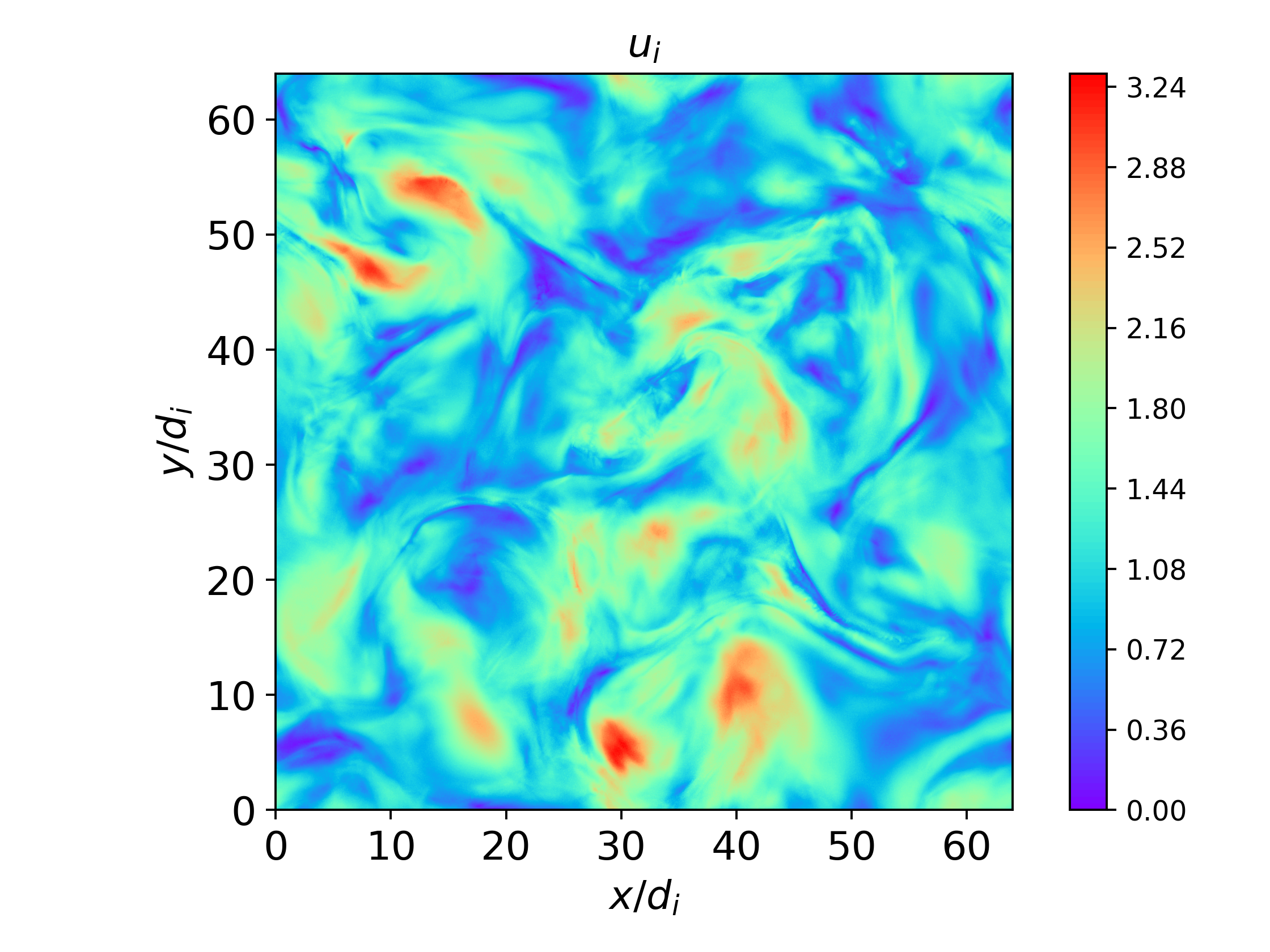}
\label{u_i}
}
\caption{Shaded contour plots of the modules of the (A) total current $J$, (B) magnetic field fluctuations $\delta B\!=\!|\textbf{B}-\textbf{B}_0|$, (C) electron velocity $u_e$ and (D) ion velocity $u_i$ at $t\!=\!650\,\Omega^{-1}_e$. $J$ and $\delta B$ are expressed in units of $J_{rms}$ and $B_0$ respectively, while the velocities $u_e$ and $u_i$ are in units of $c_A$.}
\label{F0}
\end{figure*}

\begin{figure*}[t]
\centering
\subfloat[]{
\includegraphics[width=.48\linewidth]{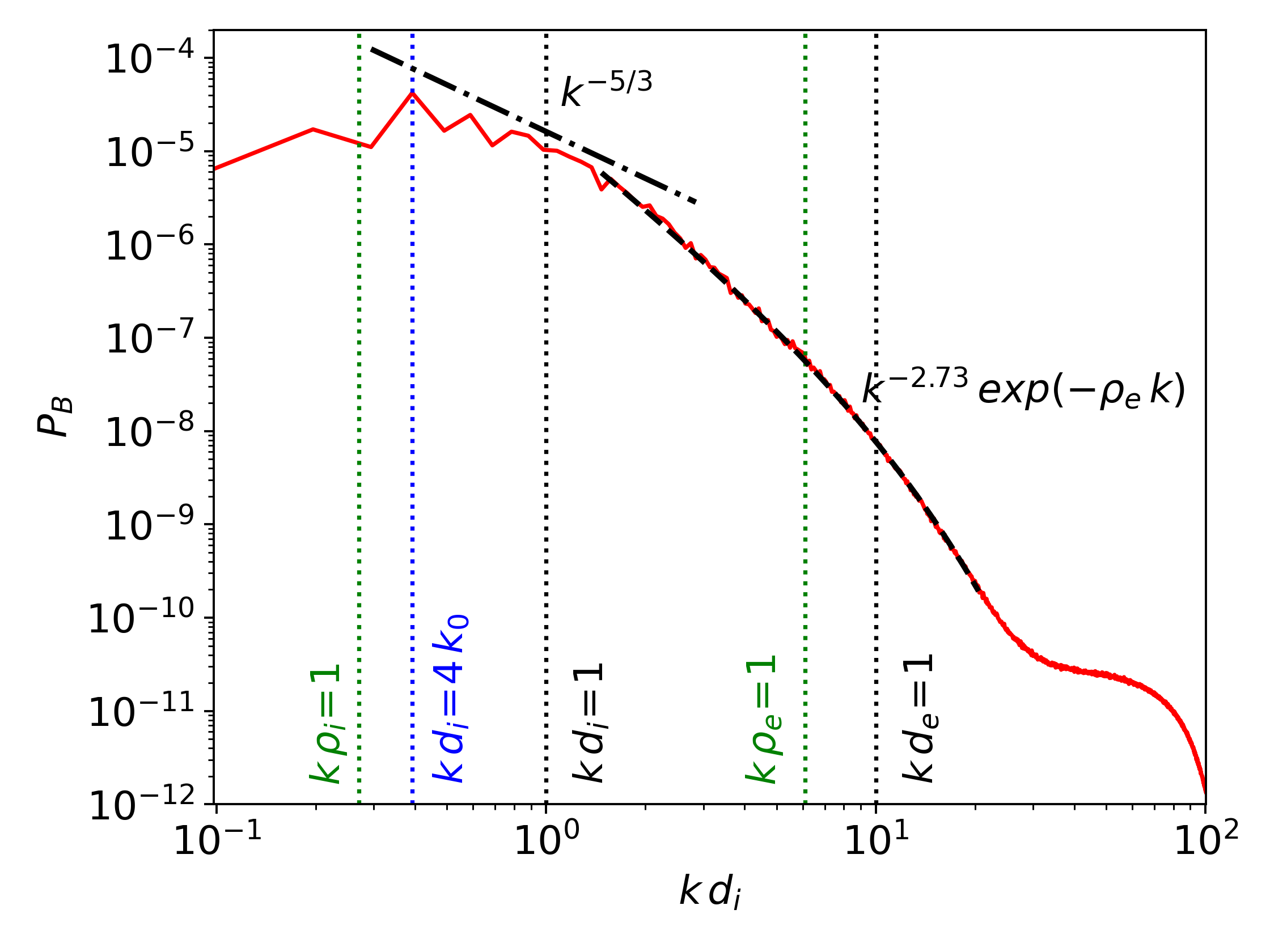}
\label{PB}
}
\hfill
\subfloat[]{
\includegraphics[width=.48\linewidth]{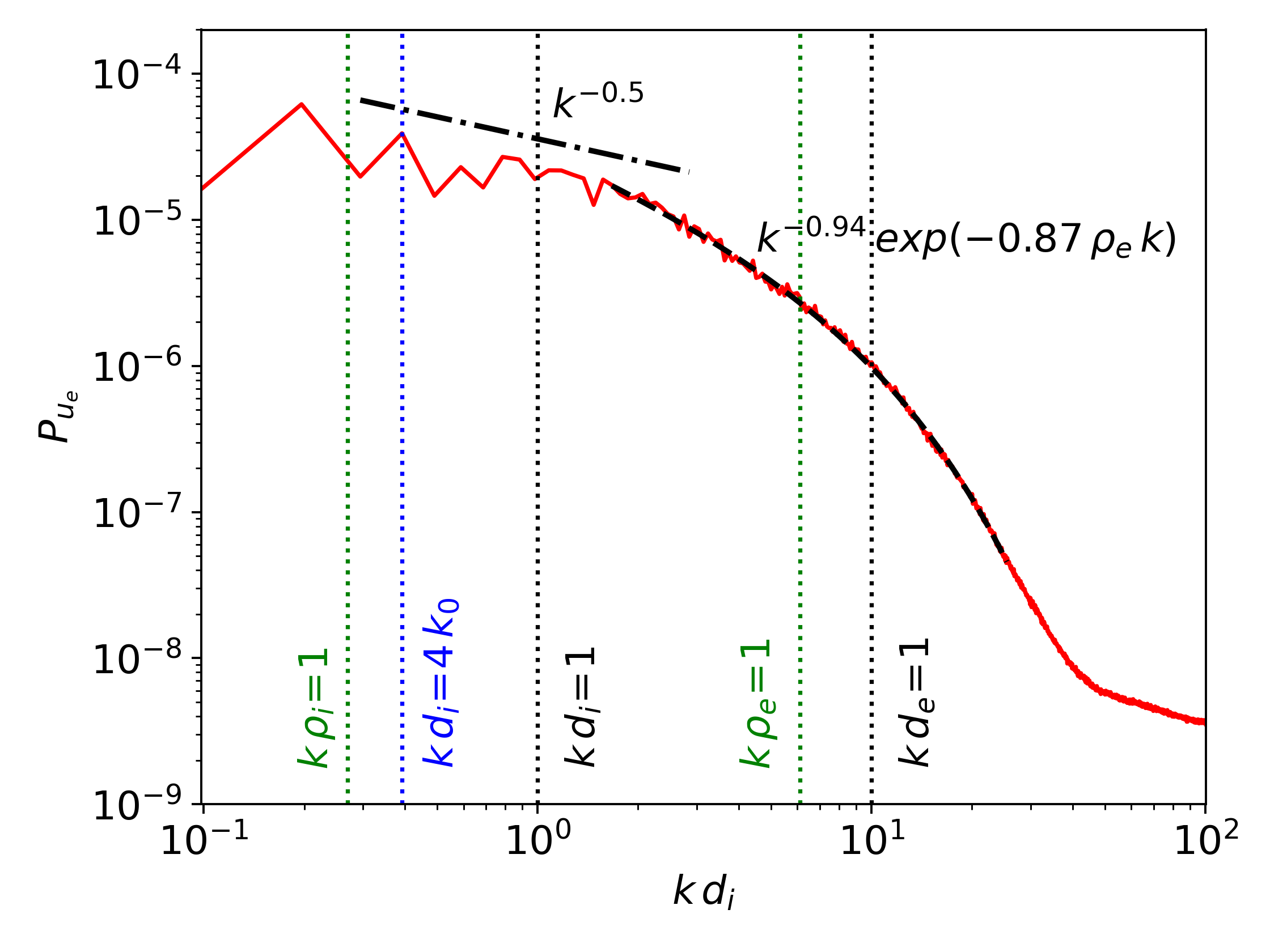}
\label{Pue}
}
\\
\subfloat[]{
\includegraphics[width=.48\linewidth]{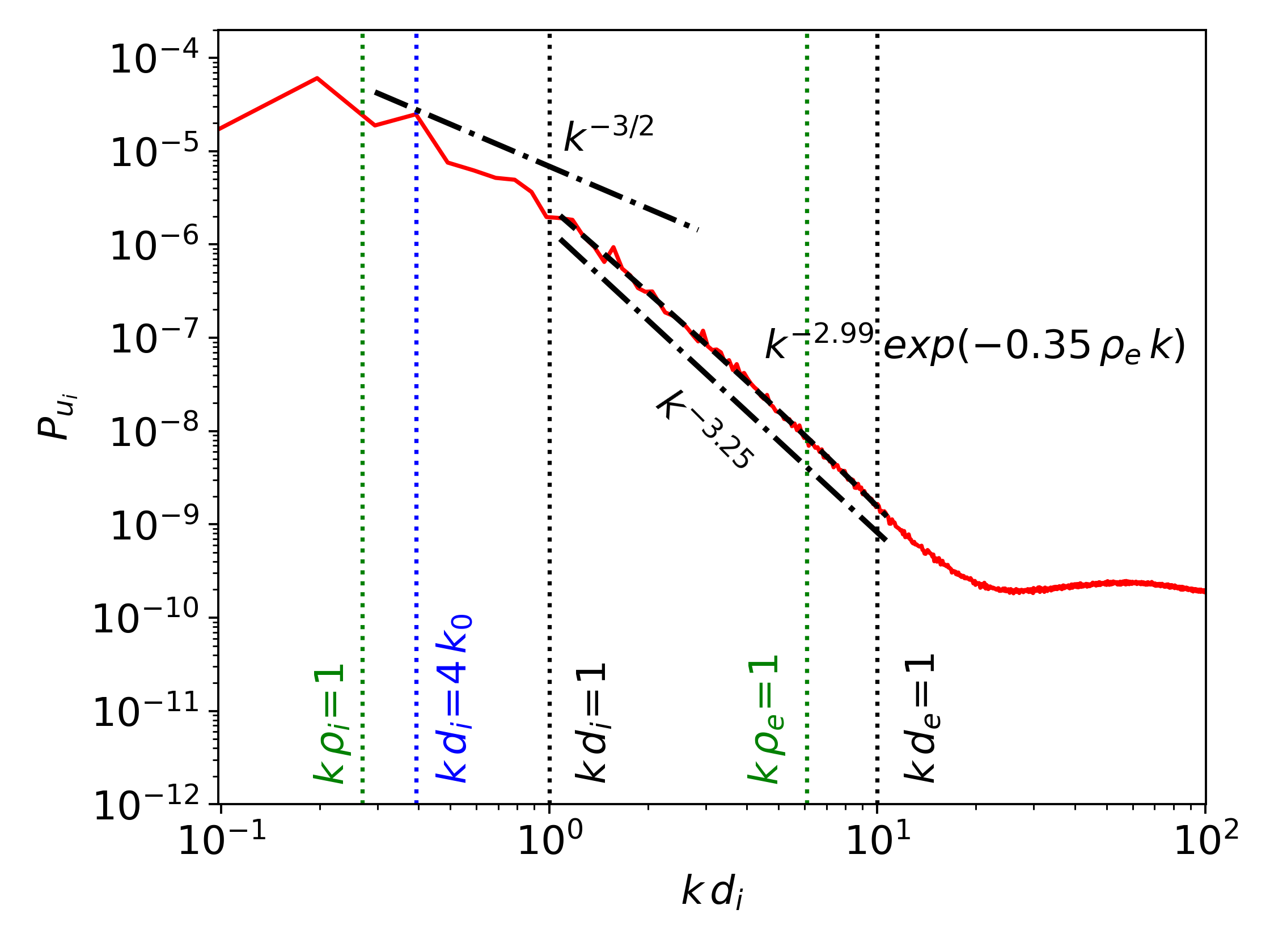}
\label{Pui}
}
\hfill
\subfloat[]{
\includegraphics[width=.48\linewidth]{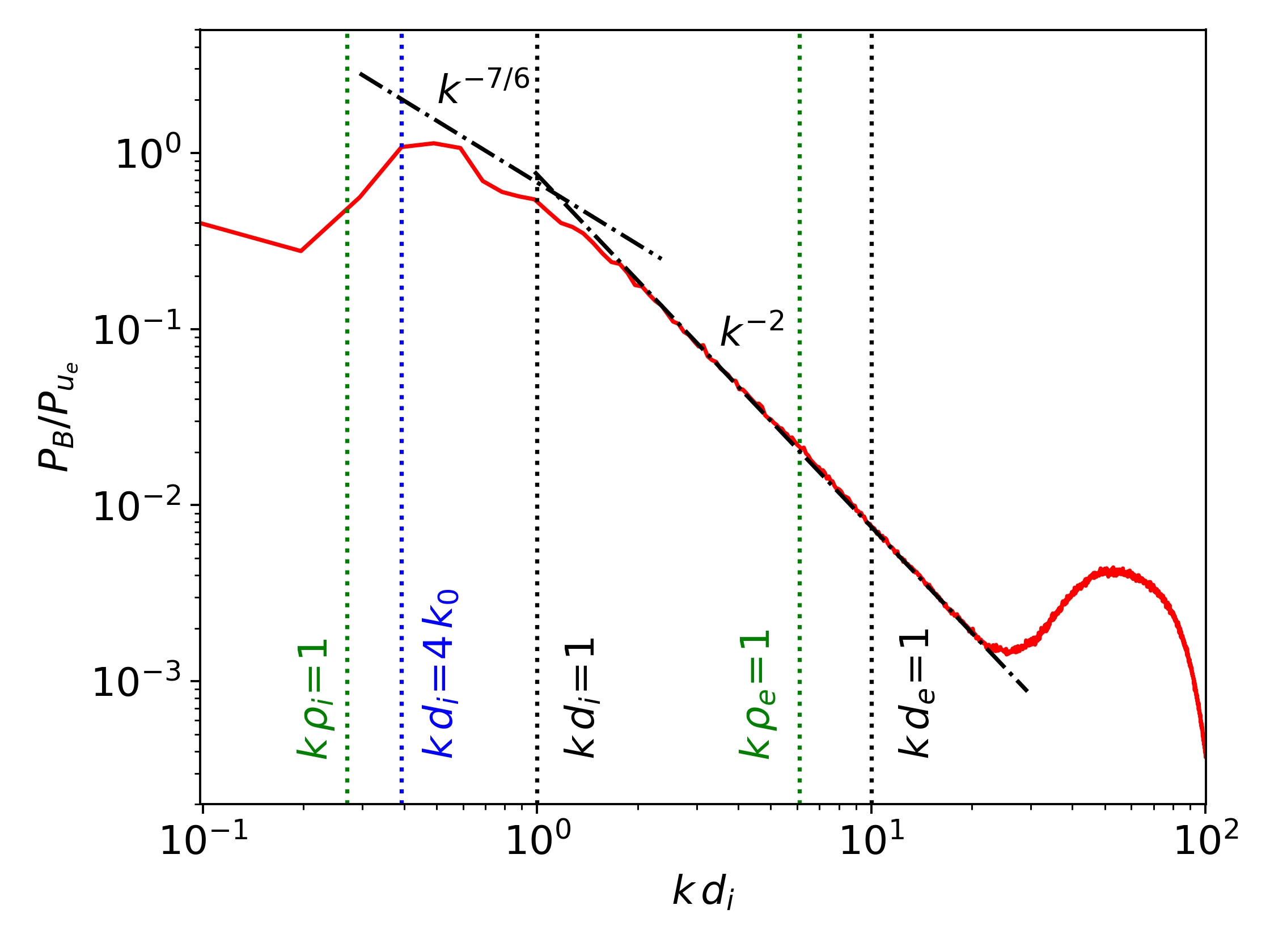}
\label{PBPue}
}
\caption{(A) Magnetic field spectrum, (B) electron velocity spectrum, (C) ion velocity spectrum and (D) ratio between the magnetic and electron velocity spectra at $t\!=\!650\,\Omega^{-1}_e$. The power laws indicated by the dash-dotted lines are given as a reference. The dashed lines indicate the fitting curves obtained using the \textit{exp} model. The vertical dotted lines indicate the injection scale $k\,d_i\!=\!4\,k_0$ and $k\,d_i\!=\!1$, $k\,d_e\!=\!1$, $k\,\rho_i\!=\!1$, $k\,\rho_e\!=\!1$ at $t\!=\!650\,\Omega^{-1}_e$.}
\label{F1}
\end{figure*}

Fig.~\ref{Energy} shows the time evolution of the energy of the system. The total energy $E_{tot}$ is well conserved, with fluctuations that are four orders of magnitudes smaller than its average value. We see that the e.m. energy $E_{e.m.}$ increases from the beginning of the simulation, reaches its maximum at about $t\!\simeq\!450\,\Omega_e^{-1}$ and then starts to decrease. The ion fluid flow energy $E_{f,i}$ monotonically decreases over time while the initially decreasing electron fluid flow energy $E_{f,e}$ starts to grow at around $t\!\simeq\!200\,\Omega_e^{-1}$, reaching its maximum at about $t\!\simeq\!500\,\Omega_e^{-1}$, after which it starts to decrease again. Both the ion and electron internal energies, $E_{th,i}$ and $E_{th,e}$ respectively, monotonically increase over time. Since we trigger the turbulence using high amplitude velocity fluctuations, we have an excess of $E_{f,i}$ with respect to $E_{e.m.}$ at the beginning of the simulation. However, as the system evolves, $E_{f,i}$ becomes smaller than $E_{e.m.}$ and when the turbulence is fully developed ($t\!>\!500\,\Omega^{-1}_e$), the residual energy $\sigma_R\!=\!(E_{f,i}-E_{e.m.})/(E_{f,i}+E_{e.m.})$ remains below $-0.2$, as typically observed both in satellite measurements \citep{chen2013residual} and numerical simulations \citep{franci2015high}. 
The root mean square ($rms$) of the total current $J_{rms}$ is represented in Fig.~\ref{Jrms}. At $t\!>\!500\,\Omega^{-1}_e$ we see that $J_{rms}$ has reached and maintains a roughly constant peak value, indicating that the turbulence is fully  developed from large to kinetic scales \citep{Mininni2009, Servidio2011, Servidio2015}. This is seen in Fig.~\ref{F0} where the shaded contour plots of the modules of the total current $J$, magnetic field fluctuations $\delta B\!=\!|\textbf{B}-\textbf{B}_0|$, electron velocity $u_e$ and ion velocity $u_i$ show a great variety of vortex-like and sheet-like structures consistent with a turbulent flow. We analyze the turbulent spectra calculated at $t\!=\!650\,\Omega^{-1}_e$, close to the maximum of $J_{rms}$, when the turbulence is fully developed. The magnetic field, electron velocity and ion velocity spectra $P_B$, $P_{u_e}$ and $P_{u_i}$ at $t\!=\!650\,\Omega^{-1}_e$ are shown in Fig.~\ref{F1}, panels (A), (B) and (C), respectively. We analyze these three spectra since they are representative of the three main forms of energy, i.e. the e.m. energy $E_{e.m.}$ and the fluid flow energies of electrons and ions, $E_{f,e}$ and $E_{f,i}$, respectively. No small scale filtering or smoothing have been applied to the spectra. With our simulation setup, the range of scales between the injection scale $k\,d_i\!=\!4\,k_0\!\simeq\!0.39$ and $k\,d_i\!\simeq\!1$ is covered only by seven wavenumbers. Nevertheless, we see that all spectra roughly follow a power law that breaks and steepens below $k\,d_i\!\simeq\!1$. In the kinetic range, that is for $k\,d_i\!>\!1$, the magnetic field and electron velocity spectra exhibit a similar behavior, showing a clear negative curvature that becomes more prominent towards $k\,d_e\!\simeq\!1$ and extends into electron scales. On the other hand, no significant curvature is observed for the ion velocity spectrum that does not extend much into electron scales. In analogy with recent studies on satellite data, we fit these spectra in the kinetic range using the \textit{exp} model proposed by \citet{Alexandrova2012} to describe the SW magnetic field spectrum at sub-ion scales. The range covered by each fit starts at around $k\,d_i\!\simeq\!1.5$, where the spectral break is observed. To decide where to stop the fits, we notice that all spectra become convex at high $k$, beyond electron scales. 
This curvature inversion is most likely caused by the intrinsic numerical noise of PIC codes whose effect is  to create small scale random fluctuations and thus a bump in the spectra at high $k$. Hence, each fit stops at the inflection point preceding the convex part of the spectrum. 

The results of the fits are shown in Fig.~\ref{F1} where the fitting curves obtained are plotted on the corresponding spectra in the kinetic range. We see that the \textit{exp} model $k^{-\alpha}\,exp(-\lambda\, k)$ fits well the magnetic field spectrum at kinetic scales, with a scaling exponent $\alpha_{B}\!\simeq\!2.73$ and a characteristic length $\lambda_{B}\!\simeq\!0.164\!\simeq\!\rho_e$ that is equal to the electron gyroradius, whose value is $\rho_e\!\simeq\!0.164$ at $t\!=\!650\,\Omega^{-1}_e$ in our simulation. 
Moreover, even the electron velocity spectrum is well described by the \textit{exp} model at kinetic scales, with a scaling exponent $\alpha_{u_e}\!\simeq\!0.94$ and a characteristic length $\lambda_{u_e}\!=\!0.142\!\simeq\!0.87\,\rho_e$ that is also of the order of the electron gyroradius $\rho_e$. Finally, for the ion velocity spectrum the \textit{exp} model fit gives a scaling exponent $\alpha_{u_i}\!=\!2.99$ and a characteristic length $\lambda_{u_i}\!=\!0.057\!\simeq\!0.35\,\rho_e$, smaller than the electron gyroradius $\rho_e$. It is important to notice that the characteristic lengths $\lambda_{B}$ and $\lambda_{u_e}$ of the magnetic field and electron velocity spectra are very similar, which means that their exponential behavior becomes dominant at about the same wavenumbers which are $k_{B} \!=\! 1/\lambda_{B}\! \simeq \!6.1$ for the magnetic field and $k_{u_e}\! = \!1/\lambda_{u_e}\! \simeq\!7$ for the electron velocity. On the other hand, the ion velocity spectrum shows a weak exponential behavior in the kinetic range and the power law scaling seems to be dominant. In fact, according to the fit, the exponential part of the ion velocity spectrum becomes relevant at $k_{u_i}\!=\!1/\lambda_{u_i}\!\simeq\!17.5$ that falls far beyond ion kinetic scales, in the interval where the spectrum is already convex and thus influenced by the numerical noise. In other words, the exponential behavior introduced by the \textit{exp} model is not relevant in the range covered by the fit in the case of the ion velocity spectrum. As a comparison, we also fit the ion velocity spectrum with a pure power law model $k^{-\beta}$ that gives a scaling exponent $\beta_{u_i}\!=\!3.25$, very close to the value $\alpha_{u_i}\!=\!2.99$ obtained with the \textit{exp} model. The $k^{-3.25}$ power law is shown in Fig.~\ref{Pui}, compared to the \textit{exp} model fit. We see that the power law alone is sufficient to describe the behavior of the ion velocity spectrum in the kinetic range and no substantial difference is observed with respect to the \textit{exp} model. This is confirmed by comparing the goodness of the fits realized with the two models, quantified as the mean square distance between the spectrum and the fitting curve. We find a goodness of $\Gamma_{power}\!=\!7.96\cdot 10^{-3}$ for the power law model, very close to the goodness $\Gamma_{exp}\!=\!6.01\cdot 10^{-3}$ found for the \textit{exp} model. Thus, we conclude that within the range of kinetic scales covered by our simulation, a power law model is the most appropriate description for the ion velocity spectrum while the magnetic field and electron velocity spectra are well described by the \textit{exp} model.

The similarities in the shapes of the magnetic field and electron velocity spectra at sub-ion scales suggest that the electrons do play a major role with respect to the ions in shaping the magnetic field spectrum at these scales, in particular by contributing to the formation of the electron scale exponential range that is instead absent in the ion velocity spectrum. This is also consistent with the fact that the ions are expected to decouple from the magnetic field dynamics at sub-ion scales \citep{Califano2020, Pyakurel2019}. In these conditions, the ion velocity fluctuations become small with respect to the electron velocity fluctuations and if the electron dynamics is mainly incompressible, from the Ampere's law it follows that $\nabla \times \textbf{B}\!\sim\!\textbf{J}\!\sim\!\textbf{u}_e$, which implies $k^2\,P_B\!\sim\!P_{u_e}$. This is indeed observed in our simulation, as shown in Fig.~\ref{PBPue} where the ratio between $P_B$ and $P_{u_e}$ follows the power law $\sim\!k^{-2}$ at sub-ion scales, over the whole range where we performed the fits, from $k\,d_i\!\simeq\!1.5$ up to about $k\,d_i\!\simeq\!20$. Hence, it is reasonable to presume that in this range only the incompressible electron dynamics continues to support the magnetic field energy cascade, thus influencing its spectral features. 

To investigate the role of the ions and of the electrons in the development of the turbulence from large to sub-ion scales, we study the filtered energy conversion channels introduced in Eq.~(\ref{T1}) and Eq.~(\ref{T2}). We want to separate the two ranges of scales above and below $k\,d_i\!\simeq\!1.5$, where the spectral breaks are observed. To do so, we consider two groups of filtered energy conversion channels that we indicate as low-pass filtered channels, describing the energy exchanges at scales $k\,d_i\!<\!1.5$, and high-pass filtered channels, accounting for the energy exchanges at scales $k\,d_i\!\geqslant\!1.5$. From now on we will refer to the scales in the range $k\,d_i\!\geqslant\!1.5$ as \enquote{sub-ion} scales while the scales in the range $k\,d_i\!<\!1.5$ will be called \enquote{large} scales (with the caveat that our \enquote{large} scales should not be interpreted as fluid scales since they are still relatively close to ion scales in our simulation, due to the limited size of the simulation domain). Following \citet{Matthaeus_2020}, the low-pass filtered e.m. work $W^<_s$ and pressure-strain interaction $PS^<_s$ are defined as:

\begin{gather}
W^<_s\!=\!\overline{\textbf{J}}_s\cdot\widehat{\textbf{E}}
\\ \notag \\
PS^<_s\!=\!-\overline{\textbf{P}}_s\!:\!\nabla\,\widehat{\textbf{u}}_s
\end{gather}
\\where the superscript $<$ indicates that these quantities describe the energy exchanges in the range $k\,d_i\!<\!1.5$. The low-pass filter $\,\overline{\cdot}\,$ is defined as in \citet{frisch1995turbulence}:

\begin{gather}
\overline{q}(\textbf{x})=\sum_{k\,d_i<1.5}Q(\textbf{k})\,exp(i\,\textbf{k}\,\textbf{x})    
\end{gather}
\\where $q(\textbf{x})$ is a generic quantity and $Q(\textbf{k})$ is its Fourier transform. The hat $\,\widehat{\cdot}\,$ indicates the density-weighted low-pass filter $\widehat{q}\!=\!\overline{q\,n}/\overline{n}$ (where $n$ is the density). The high-pass filtered energy conversion channels are obtained by subtracting the low-pass filtered channels from the corresponding unfiltered quantities. Therefore, the high-pass filtered e.m. work $W^>_s$ and pressure-strain interaction $PS^>_s$ are defined as:

\begin{gather}
W^>_s\!=\!\textbf{J}_s\cdot\textbf{E}-\overline{\textbf{J}}_s\cdot\widehat{\textbf{E}}\!=\!W_s-W_s^< 
\\ \notag \\
PS^>_s\!=\!(-\textbf{P}_s\!:\!\nabla\,\textbf{u}_s)-(-\overline{\textbf{P}}_s\!:\!\nabla\,\widehat{\textbf{u}}_s)\!=\!PS_s-PS_s^<
\end{gather}
\\where the superscript $>$ indicates that these quantities describe the energy exchanges in the range $k\,d_i\!\geqslant\!1.5$. The pressure-strain interaction can be usefully decomposed as:

\begin{gather}
PS_s\!=\!-\textbf{P}_s\!:\!\nabla\,\textbf{u}_s\!=\!-p_s\nabla\cdot\textbf{u}_s-\pmb{\Pi}_s\!:\!\textbf{D}_s\!=\!-P\theta_s+PiD_s \label{PSdec}
\end{gather}
\\with $p_s\!=\!tr(\textbf{P}_s)/3$, $\pmb{\Pi}_s\!=\!\textbf{P}_s-p_s\,\textbf{I}$ and $\textbf{D}_s\!=\![(\nabla\,\textbf{u}_s)+(\nabla\,\textbf{u}_s)^T]/2-(\nabla\cdot\textbf{u}_s/3)\,\textbf{I}$ (where $tr(\cdot)$ is the trace operation, $\textbf{I}$ is the identity matrix and the superscript $T$ indicates the transpose operation). The first term on the right-hand side of Eq.~(\ref{PSdec}), called the \enquote{P-$\theta$} interaction, describes the increase of internal energy related to isotropic compression and expansion while the second term, called the \enquote{Pi-D} interaction, accounts for the increase of internal energy caused by volume-preserving anisotropic deformations and can be interpreted as a collisionless \enquote{viscosity} \citep{DelSarto2017,Yang2017_2,Yang2017_1,Matthaeus_2020}. This decomposition separates the compressible dynamics, described by the P-$\theta$ interaction, from the incompressible dynamics, described by the Pi-D interaction. Therefore, we analyze the low-pass filtered and high-pass filtered P-$\theta$ and Pi-D interactions of both ions and electrons to check whether the sub-ion scale dynamics is actually incompressible as suggested by the relation between $P_B$ and $P_{u_e}$ shown in Fig.~\ref{PBPue}. The transfer of energy between the two ranges of scales above and below $k\,d_i\!\simeq\!1.5$ is described in terms of the cross-scale energy fluxes $\Pi^{uu}_s$ and $\Pi^{bb}_s$, defined as (see \citet{Matthaeus_2020}):

\begin{gather}
\Pi^{uu}_s\!=\!-m_s\,\overline{n}_s\left(\,\reallywidehat{\textbf{u}_s\textbf{u}_s}-\widehat{\textbf{u}}_s\widehat{\textbf{u}}_s\,\right)\!:\!\nabla\,\widehat{\textbf{u}}_s-\frac{q_s}{c}\,\overline{n}_s\left(\,\reallywidehat{\textbf{u}_s\times\textbf{B}}\,\right)\cdot\widehat{\textbf{u}}_s
\\ \notag \\
\Pi^{bb}_s\!=\!\overline{\textbf{J}}_s\cdot\left(\,\overline{\textbf{E}}-\widehat{\textbf{E}}\,\right)
\end{gather}
\\where $c$ is the speed of light. The cross-scale flux $\Pi^{uu}_s$ represents the fluid flow energy transferred from the range $k\,d_i\!<\!1.5$ into the range $k\,d_i\!\geqslant\!1.5$. Similarly, the cross-scale flux $\Pi^{bb}_s$ accounts for the e.m. energy transferred from the range $k\,d_i\!<\!1.5$ into the range $k\,d_i\!\geqslant\!1.5$. Since we are interested in the global energy balance, we analyze the box-averaged energy conversion channels, indicating the average operation with the symbol $\langle \cdot \rangle$. 

\begin{figure*}[t]
\centering
\subfloat[]{
\includegraphics[width=.48\linewidth]{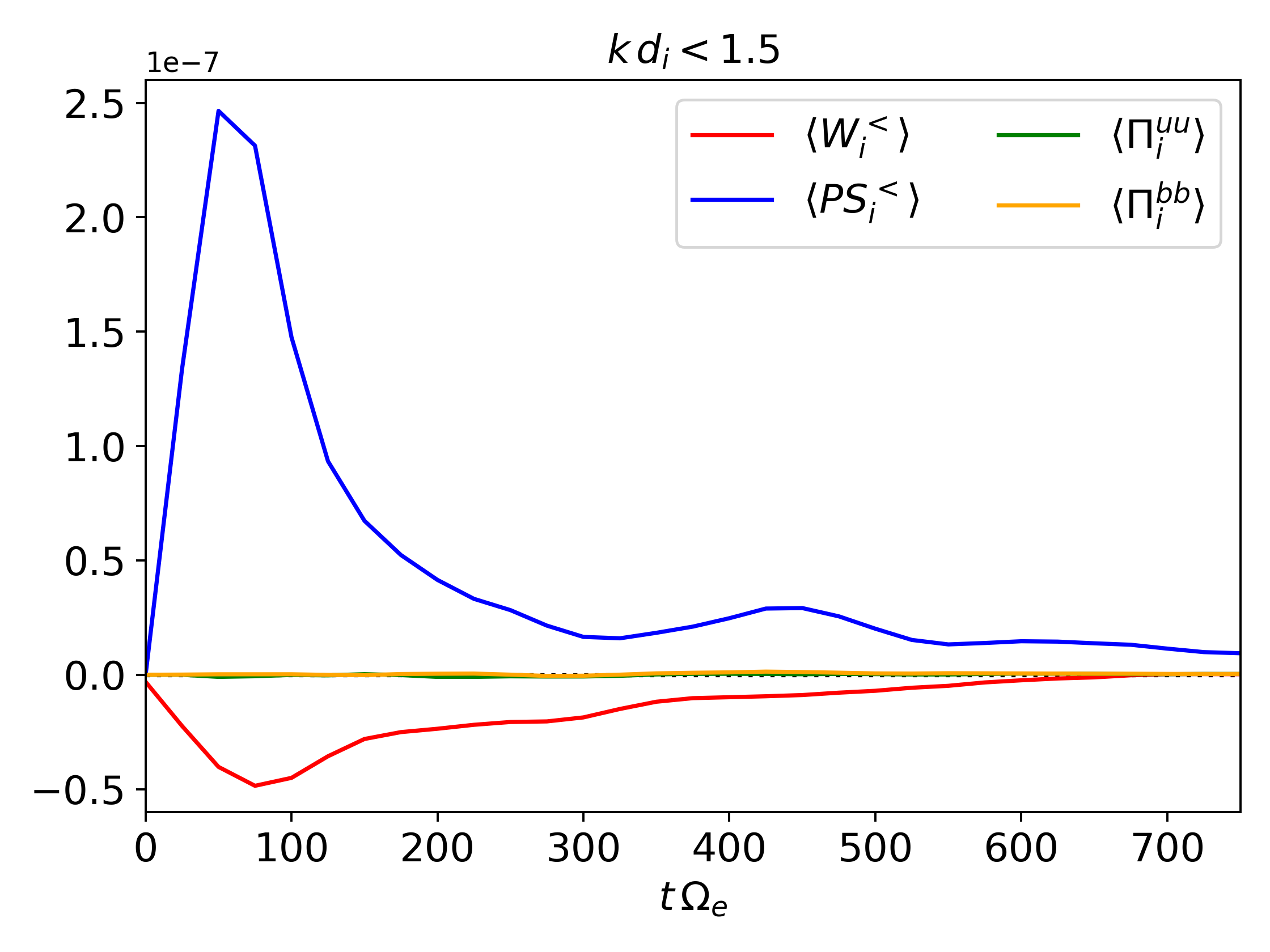} 
\label{ECC_i_<}
}
\hfill
\subfloat[]{
\includegraphics[width=.48\linewidth]{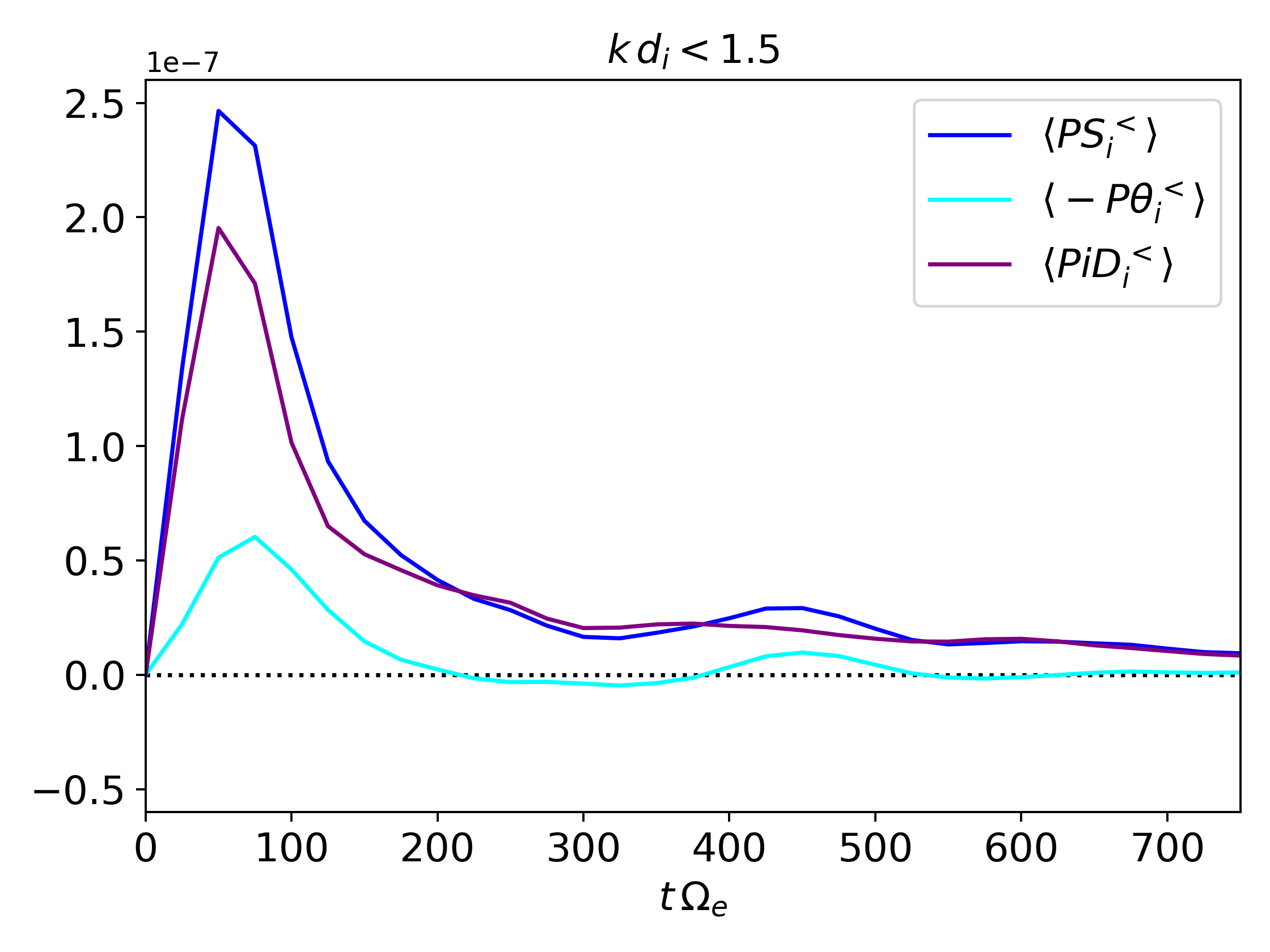}
\label{PS_i_<}
}
\\
\subfloat[]{
\includegraphics[width=.48\linewidth]{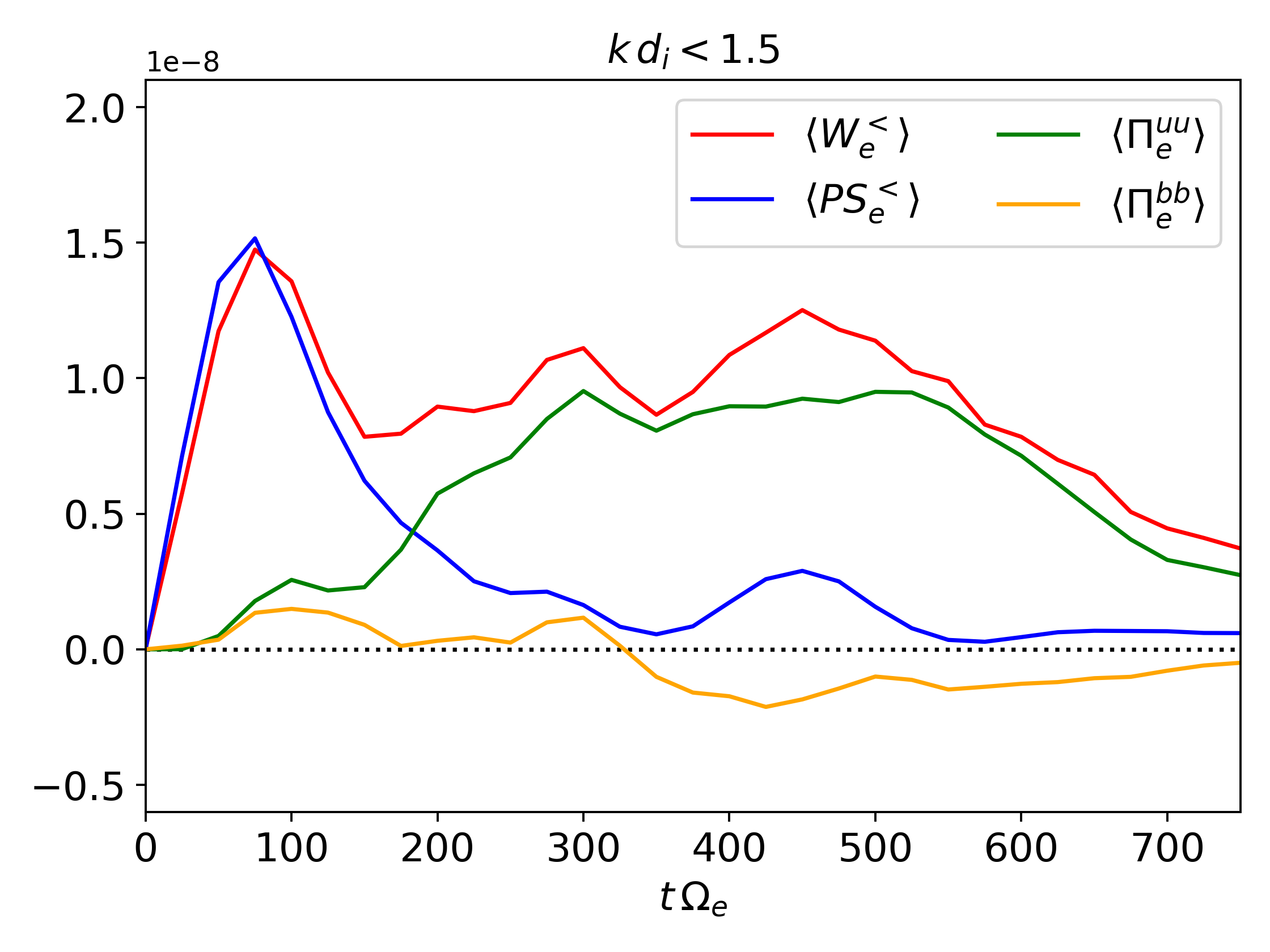}
\label{ECC_e_<}
}
\hfill
\subfloat[]{
\includegraphics[width=.48\linewidth]{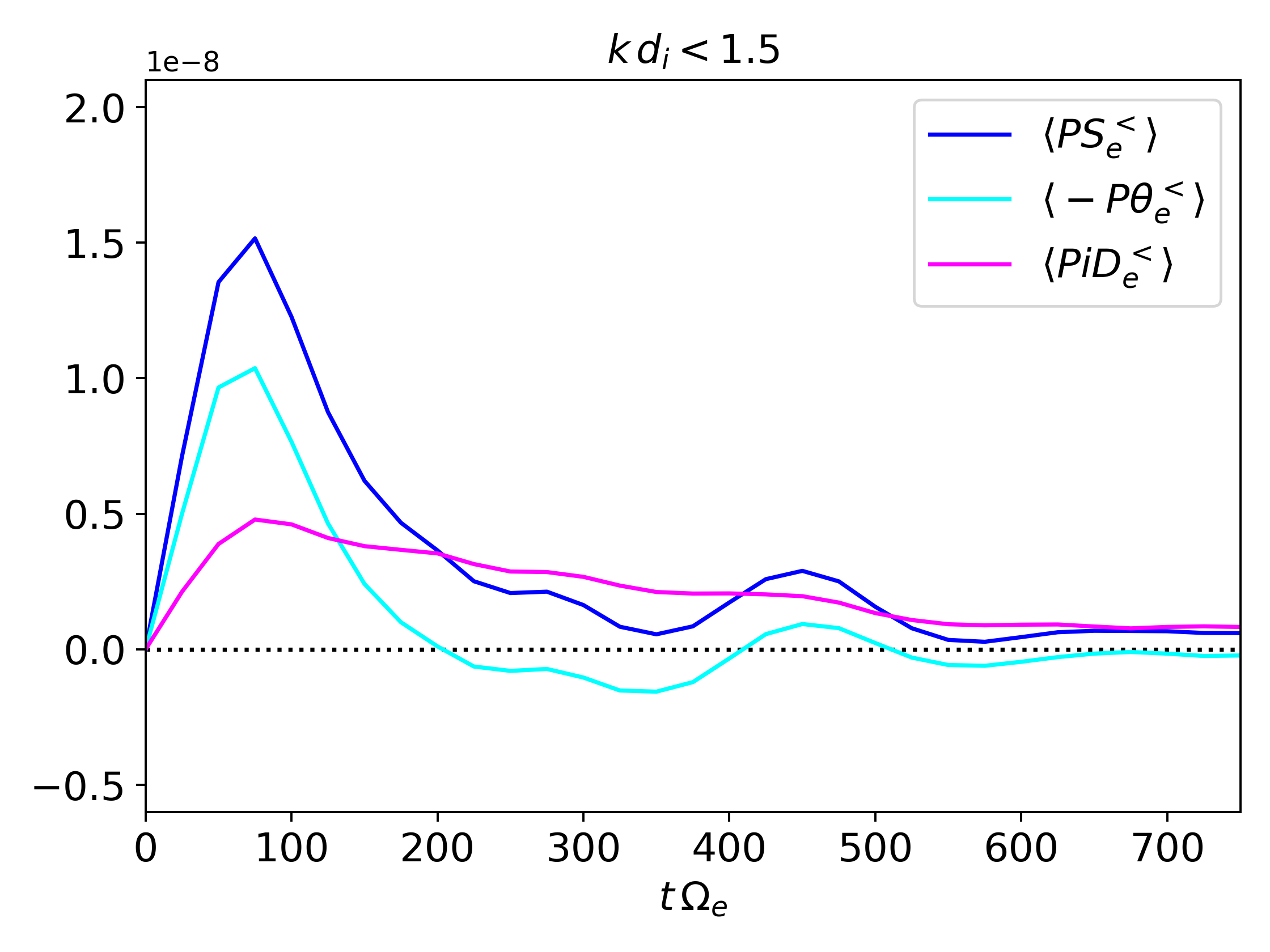}
\label{PS_e_<}
}
\caption{Time evolution of the box-averaged low-pass filtered energy conversion channels of (A) ions and (C) electrons at scales $k\,d_i\!<\!1.5$. Decomposition of the low-pass filtered pressure-strain interaction for (B) ions and (D) electrons. The horizontal black dotted line is centered at zero.}
\label{F2}
\end{figure*}

Fig.~\ref{F2} shows the time evolution of the box-averaged low-pass filtered ion and electron energy conversion channels. In Fig.~\ref{ECC_i_<} we see that the dominant channels determining the energy balance of the ions at scales $k\,d_i\!<\!1.5$ are the e.m. work $\langle W^<_i \rangle$ and the pressure-strain interaction $\langle PS^<_i \rangle$, while the cross-scale energy fluxes $\langle \Pi^{uu}_i \rangle$ and $\langle \Pi^{bb}_i \rangle$ are way smaller in magnitude. $\langle W^<_i \rangle$ is negative from the beginning of the simulation and approaches zero after the turbulence is fully developed ($t\!>\!500\,\Omega^{-1}_e$), meaning that the ions are giving energy to the e.m. field at scales $k\,d_i\!<\!1.5$. On the other hand, $\langle PS^<_i \rangle$ is positive during the whole simulation, which means that the ion fluid flow energy is being converted also into internal energy at large scales. The decomposition of $\langle PS^<_i \rangle$ in Fig.~\ref{PS_i_<} shows that most of the ion heating at large scales results from an incompressible dynamics since the main contribution to $\langle PS^<_i \rangle$ comes from the \enquote{Pi-D} interaction $\langle PiD^<_i \rangle$. Differently from the ions, in Fig.~\ref{ECC_e_<} we see that in the case of the electrons the cross-scale fluxes $\langle \Pi^{uu}_e \rangle$ and $\langle \Pi^{bb}_e \rangle$ are of the same order of $\langle W^<_e \rangle$ and $\langle PS^<_e \rangle$ at scales $k\,d_i\!<\!1.5$. $\langle W^<_e \rangle$ is positive during the whole simulation, meaning that the electrons are taking energy from the e.m. field at large scales. $\langle PS^<_e \rangle$ and $\langle \Pi^{uu}_e \rangle$ are also positive, which means that the fluid flow energy gained via the e.m. work $\langle W^<_e \rangle$ is partially converted into internal energy by the $\langle PS^<_e \rangle$ interaction while another consistent fraction is transferred to sub-ion scales by the cross-scale flux of fluid flow energy $\langle \Pi^{uu}_e \rangle$. The electron cross-scale flux of e.m. energy $\langle \Pi^{bb}_e \rangle$ is positive from the beginning of the simulations and turns negative at about $t\!\simeq\!330\,\Omega_e^{-1}$. This means that the electrons are initially supporting the transfer of e.m. energy from large to sub-ion scales, while for $t\!>\!330\,\Omega_e^{-1}$ this flux of energy reverts and the e.m. energy at sub-ion scales is transferred to scales $k\,d_i\!<\!1.5$. The $\langle PS^<_e \rangle$  decomposition in Fig.~\ref{PS_e_<} shows that for $t\!<\!135\,\Omega_e^{-1}$ the large scale electron dynamics is mainly compressible since most of the pressure-strain interaction is given by $\langle P\theta^<_e \rangle$. However, for $t\!>\!135\,\Omega_e^{-1}$ the $\langle PiD^<_e \rangle$ interaction becomes dominant with respect to $\langle P\theta^<_e \rangle$ and the electron dynamics becomes mainly incompressible. We notice that $\langle W^<_i \rangle$, $\langle PS^<_i \rangle$, $\langle W^<_e \rangle$ and $\langle PS^<_e \rangle$ have a peak at about $t\!\simeq\!75\,\Omega_e^{-1}$, that quickly decreases for $t\!>\!100\,\Omega_e^{-1}$. 
\begin{figure*}[t]
\centering
\subfloat[]{
\includegraphics[width=.48\linewidth]{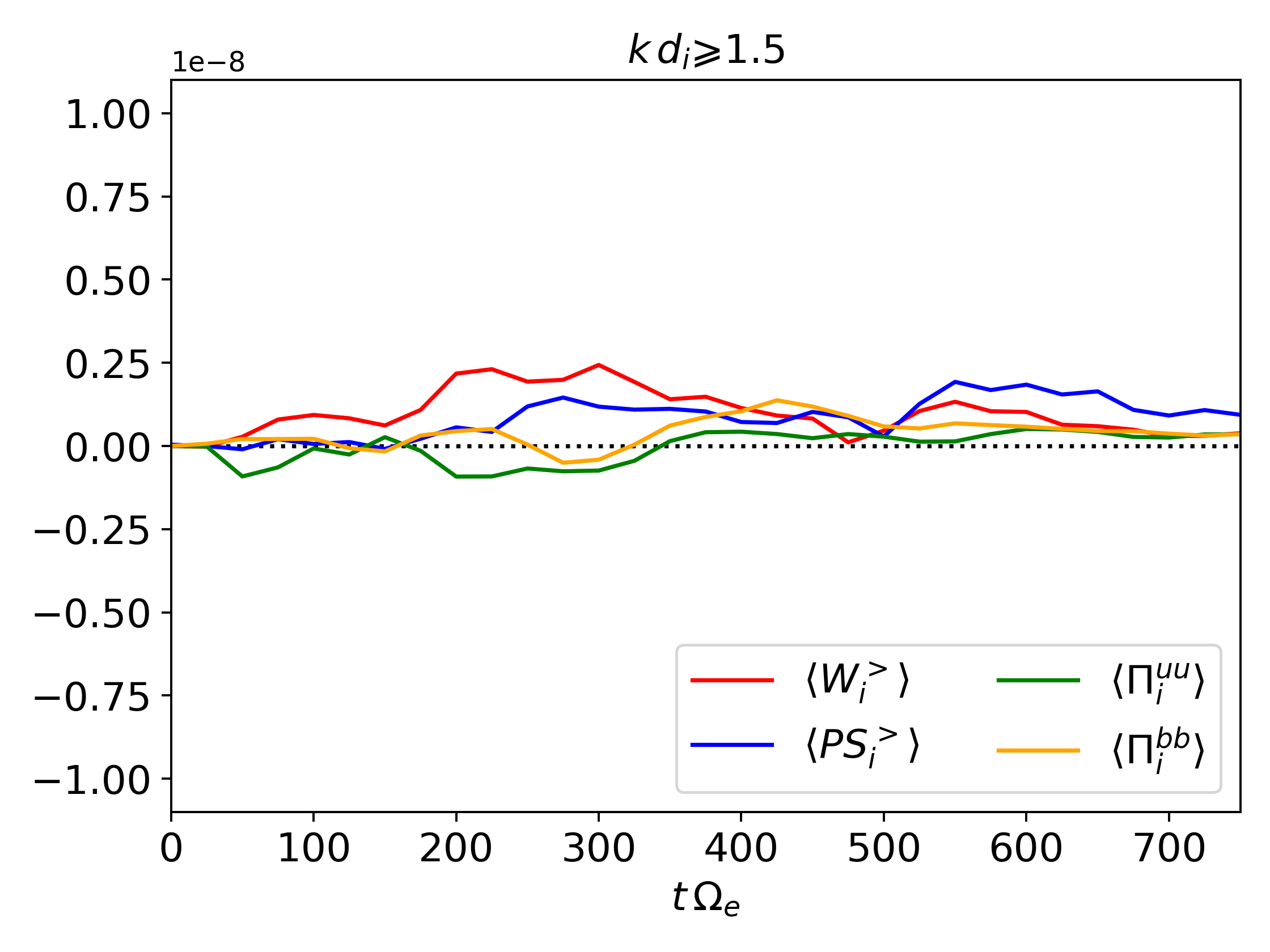} 
\label{ECC_i_>}
}
\hfill
\subfloat[]{
\includegraphics[width=.48\linewidth]{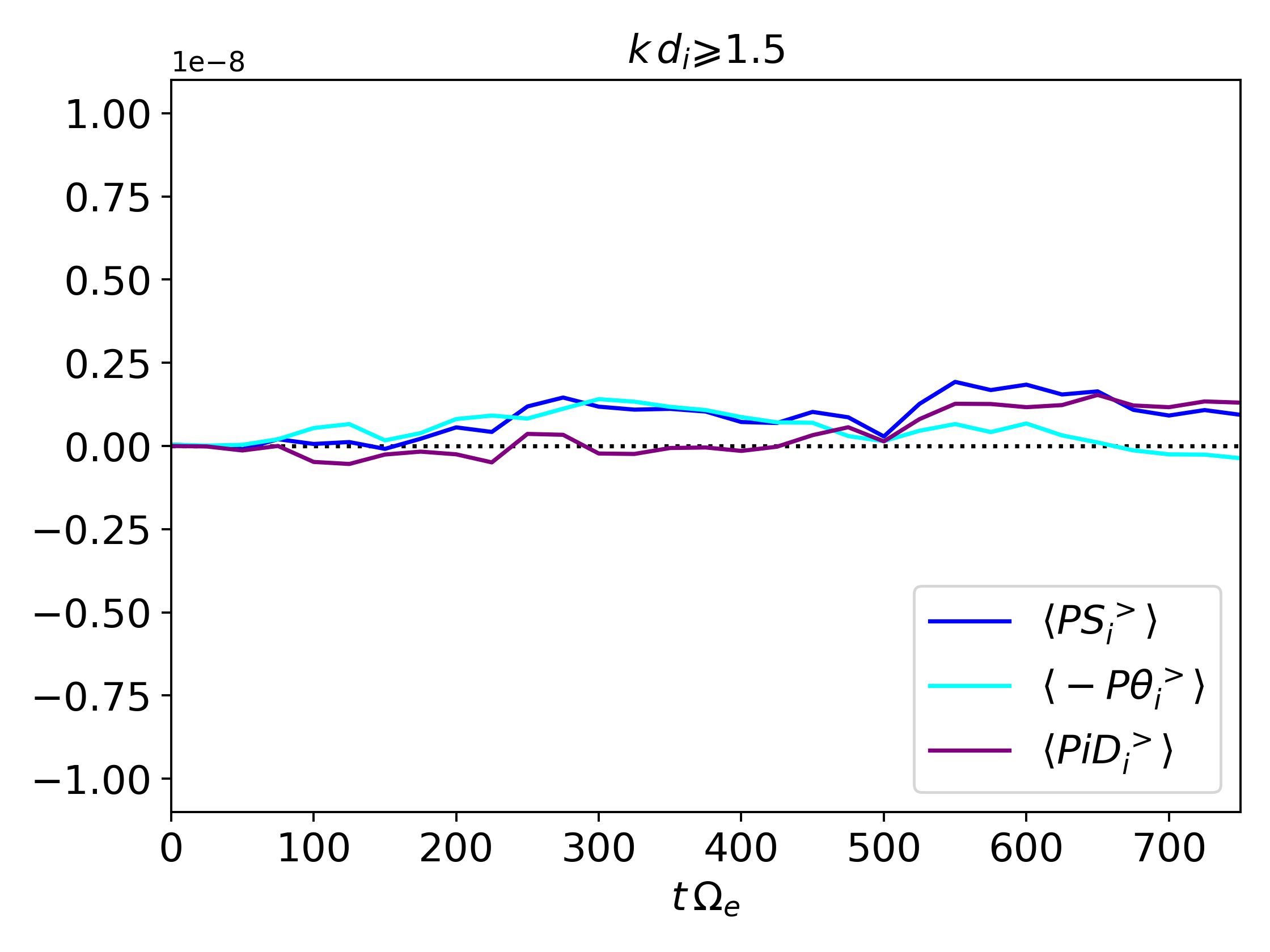}
\label{PS_i_>}
}
\\
\subfloat[]{
\includegraphics[width=.48\linewidth]{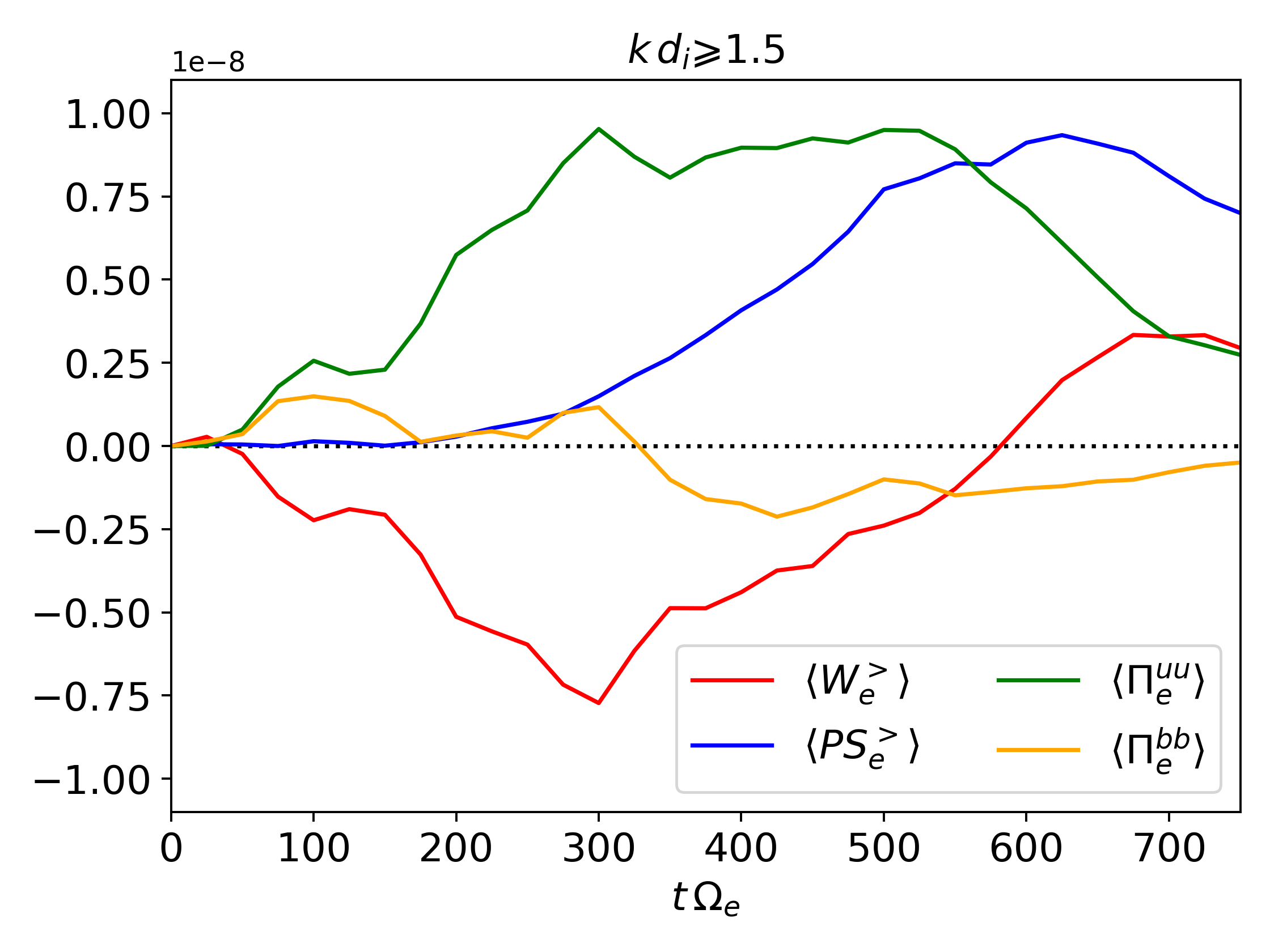}
\label{ECC_e_>}
}
\hfill
\subfloat[]{
\includegraphics[width=.48\linewidth]{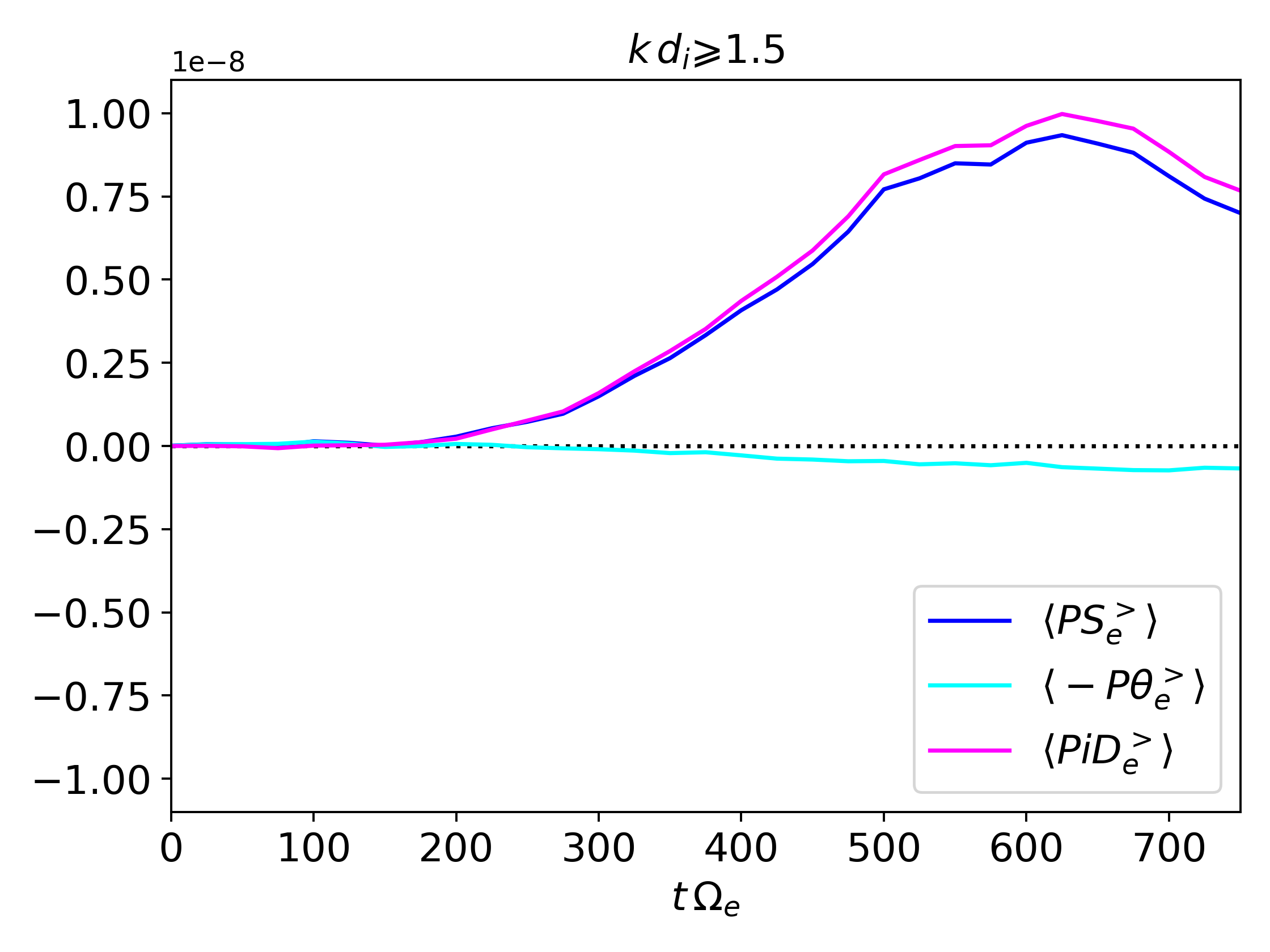}
\label{PS_e_>}
}
\caption{Time evolution of the box-averaged high-pass filtered energy conversion channels of (A) ions and (C) electrons at scales $k\,d_i\!\geqslant\!1.5$. Decomposition of the high-pass filtered pressure-strain interaction for (B) ions and (D) electrons. The horizontal black dotted line is centered at zero.}
\label{F3}
\end{figure*}
The presence of this peak is caused by the strong compression and deformation the initial high-amplitude magnetic and velocity fluctuations undergo at the beginning of the simulation. This initial violent dynamics acts as a driver for the formation of current-sheet structures but as soon as they start to form, the peaks quickly settle down.

Fig.~\ref{F3} shows the time evolution of the box-averaged high-pass filtered ion and electron energy conversion channels. In Fig.~\ref{ECC_i_>} we see that the main source of energy for the ions at $k\,d_i\!\geqslant\!1.5$ is the e.m. work $\langle W^>_i \rangle$, which remains positive during the whole simulation. This means that the ions are constantly taking energy from the e.m. field at sub-ion scales. Part of this energy is converted into internal energy via the pressure-strain interaction $\langle PS^>_i \rangle$ that is also positive during the whole simulation. Only a very small fraction of the large scale ion fluid flow energy is transferred to $k\,d_i\!\geqslant\!1.5$, as shown by the cross-scale flux $\langle \Pi^{uu}_i \rangle$ that stays negative during the first half of the simulations and turns slightly positive only at about $t\!\simeq\!330\,\Omega_e^{-1}$. The contribution of the ions to the transport of e.m. energy from large scales to $k\,d_i\!\geqslant\!1.5$ is also weak. The cross-scale flux $\langle \Pi^{bb}_i \rangle$ initially fluctuates around zero, indicating the absence of a precise direction for the flow of energy between large and sub-ion scales, and finally becomes slightly positive at about $t\!\simeq\!330\,\Omega_e^{-1}$. The $\langle PS^>_i \rangle$ decomposition in Fig.~\ref{PS_i_>} shows that up to about $t\!\simeq\!500\,\Omega_e^{-1}$, the main contribution to the pressure-strain interaction comes from $\langle P\theta^>_i \rangle$ while only at later times $\langle PiD^>_i \rangle$ becomes dominant. As for the electrons, the main source of energy at $k\,d_i\!\geqslant\!1.5$ is the cross-scale flux of fluid flow energy $\langle \Pi^{uu}_e \rangle$ that remains positive throughout the simulation. The pressure-strain interaction $\langle PS^>_e \rangle$ is also positive and rapidly grows in time, reaching a roughly constant value at about $t\!\simeq\!500\,\Omega_e^{-1}$. The e.m. work $\langle W^>_e \rangle$ is negative from the beginning of the simulation and changes sign at about $t\!\simeq\!570\,\Omega_e^{-1}$, after which it remains positive. This change of sign can be understood by comparing $\langle W^>_e \rangle$ with $\langle PS^>_e \rangle$ and $\langle \Pi^{uu}_e \rangle$. Among these three channels, $\langle \Pi^{uu}_e \rangle$ is the largest one up to $t\!\simeq\!570\,\Omega_e^{-1}$ and it is the only term providing energy to the electrons at sub-ion scales. On the other hand, $\langle PS^>_e \rangle$ is positive and $\langle W^>_e \rangle$ is negative for $t\!<\!570\,\Omega_e^{-1}$, which means that the energy delivered by $\langle \Pi^{uu}_e \rangle$ is converted both into internal and e.m. energy, by $\langle PS^>_e \rangle$ and $\langle W^>_e \rangle$ respectively. However, around $t\!\simeq\!570\,\Omega_e^{-1}$ we see that $\langle PS^>_e \rangle$ becomes larger than $\langle \Pi^{uu}_e \rangle$. This implies that dissipation at sub-ion scales becomes more efficient than the transfer of energy coming from large scales since $\langle PS^>_e \rangle$ starts to convert into internal energy more fluid flow energy than the amount provided by $\langle \Pi^{uu}_e \rangle$. At this point the electrons are losing fluid flow energy and to compensate for this loss they start to take energy back from the e.m. field, making $\langle W^>_e \rangle$ become positive. Therefore, it is possible to distinguish two phases in the evolution of the electrons at sub-ion scales: a first phase for $t\!<\!570\,\Omega_e^{-1}$, during which the electron dynamics is driven by the cross-scale flux of energy $\langle \Pi^{uu}_e \rangle$, and a second phase for $t\!>\!570\,\Omega_e^{-1}$, dominated by the pressure-strain interaction $\langle PS^>_e \rangle$ that takes over in guiding the electron dynamics. Finally, the pressure-strain decomposition in Fig.~\ref{PS_e_>} shows that $\langle PS^>_e \rangle$ is almost entirely determined by the $\langle PiD^>_e \rangle$, with a very small negative contribution coming from $\langle P\theta^>_e \rangle$, indicating a weak expansion. This means that the electron dynamics at $k\,d_i\!\geqslant\!1.5$ is basically incompressible, consistently with the spectral features of the ratio $P_B/P_{u_e}$ in Fig.~\ref{PBPue}. 

\begin{figure}[t]
\centering
\includegraphics[width=.9\linewidth]{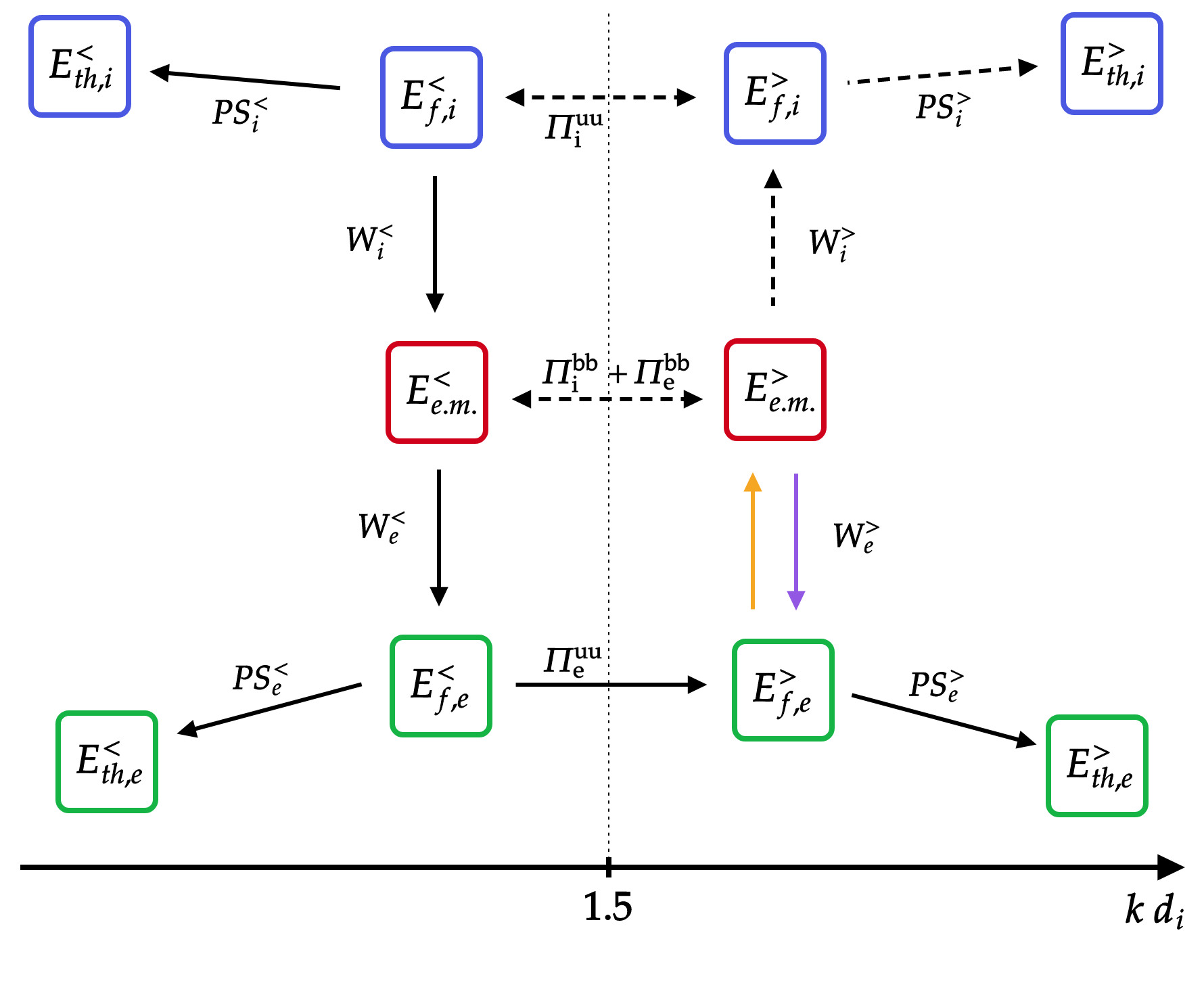} 
\caption{Diagram of the global energy balance between large and sub-ion scales in our simulation. The superscripts $<$ and $>$ indicate quantities at scales $k\,d_i\!<\!1.5$ and $k\,d_i\!\geqslant\!1.5$ respectively. The arrows indicate the direction in which the energy is transferred by the corresponding channels (the double arrows indicate no preferred direction). Solid arrows indicate strong channels while weak channels are represented by dashed arrows. The two colors used for $W_e^>$ indicate $t\!<\!570\,\Omega_e^{-1}$ (orange) and $t\!>\!570\,\Omega_e^{-1}$ (purple).}
\label{EnergyBalance}
\end{figure}

By comparing the low-pass and the high-pass filtered energy conversion channels, we can finally get an overview of the global energy balance to understand how the e.m. energy is transferred from large scales to sub-ion scales in our simulation. Fig.~\ref{EnergyBalance} shows a diagram with all the different forms of energy at $k\,d_i\!<\!1.5$ and $k\,d_i\!\geqslant\!1.5$ (identified by the superscripts $<$ and $>$ respectively), together with the energy conversion channels linking them (indicated by the arrows). From this diagram it is possible to identify two paths that the e.m. energy can follow in order to be transferred from $k\,d_i\!<\!1.5$ to $k\,d_i\!\geqslant\!1.5$. The first path is the direct scale-to-scale transfer mediated by the total cross-scale flux of e.m. energy, indicated by $\Pi^{bb}_i\!+\!\Pi^{bb}_e$ in Fig.~\ref{EnergyBalance}, that directly converts $E_{e.m.}^<$ into $E_{e.m.}^>$. As discussed above, the cross-scale transfer of e.m. energy is not particularly efficient and does not have a preferred direction (its sign changes over time). The flux $\Pi^{bb}_i$ associated to the ions becomes positive only during the second half of the simulation and its contribution to the energy balance is weak, as seen in Fig.~\ref{ECC_i_>}. On the other hand, the flux $\Pi^{bb}_e$ associated to the electrons is positive during the first half of the simulation but then it even turns negative, meaning that the e.m. energy is actually flowing towards large scales though this channel, as seen in Fig.~\ref{ECC_e_>}. Therefore, $\Pi^{bb}_i\!+\!\Pi^{bb}_e$ is not the main channel responsible for the transfer of e.m. energy from large to sub-ion scales. The second path able to connect $E_{e.m.}^<$ to $E_{e.m.}^>$ involves an indirect transfer of energy driven by the electrons and articulated in three steps: in the first step the e.m. energy $E_{e.m.}^<$ is converted into the fluid flow energy $E_{f,e}^<$ via the e.m. work $W_e^<$ at scales $k\,d_i\!<\!1.5$; in the second step $E_{f,e}^<$ is transferred to $E_{f,e}^>$ at sub-ion scales via the cross-scale flux $\Pi^{uu}_e$; in the last step the fluid flow energy $E_{f,e}^>$ is finally converted into the e.m. energy $E_{e.m.}^>$ via the e.m. work $W_e^>$ at scales $k\,d_i\!\geqslant\!1.5$. From Fig.~\ref{ECC_e_<} and Fig.~\ref{ECC_e_>} we see that $W_e^<$, $\Pi_e^{uu}$ and $W_e^>$ transfer way more energy than $\Pi^{bb}_i\!+\!\Pi^{bb}_e$, meaning that this second path is the most efficient one in transporting the e.m. energy from large to sub-ion scales. As previously discussed, we specify that the second path is accessible only for $t\!<\!570\,\Omega_e^{-1}$ since the sign of $W_e^>$ depends on the relative strength of the cross-scale flux $\Pi_e^{uu}$ with respect to the pressure-strain interaction $PS_e^>$. In particular, this mechanisms ceases to act when $PS_e^>$ exceeds $\Pi_e^{uu}$ at $t\!\simeq\!570\,\Omega_e^{-1}$. 
At that point $W_e^>$ changes sign and $E_{e.m.}^>$ starts to be transferred back to $E_{f,e}^>$ that is in turn converted into $E_{th,e}^>$ and thus dissipated. The transition from the $\Pi_e^{uu}$-dominated phase to the $PS_e^>$-dominated phase is highlighted in Fig.~\ref{EnergyBalance} by the presence of the two arrows associated to $W_e^>$ (the orange one is referred to $t\!<\!570\,\Omega_e^{-1}$ while the purple one is referred to $t\!>\!570\,\Omega_e^{-1}$). We finally observe that, differently from the electrons, the ions do not participate to the transfer of e.m. energy from large to sub-ion scales. On the contrary, since the cross-scale flux $\Pi_i^{uu}$ is weak, $E_{e.m.}^>$ represents the main source of energy for the ions at sub-ion scales, from which they draw upon via $W_i^>$. In other words, this analysis shows that the development of the e.m. field dynamics at sub-ion scales is indeed guided mainly by the electrons that support the transfer of e.m. energy from large to sub-ion scales up to $t\!\simeq\!570\,\Omega_e^{-1}$, when the system has reached a fully developed turbulent state.

\section{Discussion and conclusions}

In this work we used a fully kinetic energy conserving PIC simulation of 2D freely decaying plasma turbulence to study the spectral properties and the energy exchanges characterizing the turbulence at sub-ion scales. 

We found that the magnetic field spectrum is accurately described by the \textit{exp} model $k^{-\alpha}\,exp(-\lambda\, k)$ at sub-ion scales, with a scaling exponent $\alpha_B\!\simeq\!2.73$ and a characteristic length $\lambda_B\!\simeq\!\rho_e$, of the order of the electron gyroradius. To our knowledge, the \textit{exp} model has been tested on SW data but not in the magnetosheath, whose conditions are closer to those of our simulation. Therefore, we do not have a direct comparison with observations at the moment. Nonetheless, we have found that the functional form proposed by \citet{Alexandrova2012} works also for the magnetic field spectrum produced by our simulation, with parameters consistent with SW observations in the sense that we obtained a scaling exponent close to the SW value $-2.8$ and a characteristic length associated to the exponential range of the order of the electron gyroradius $\rho_e$ (even though the actual value of $\rho_e$ is different in the SW with respect to our simulation since it depends on the electron plasma beta and on other parameters). Furthermore, a couple of remarks regarding the range over which the fit has been performed are needed. Typically, SW observations show that in correspondence of the ion scale spectral break the slope of the magnetic spectrum varies between $-2$ and $-4$, reaching a roughly universal $-2.8$ spectral index at smaller scales. Indeed, in the works of \cite{Alexandrova2012,Alexandrova2021} the \textit{exp} model is used to fit the magnetic spectrum over a range that starts below the ion scale spectral break, where the $\sim\!k^{-2.8}$ scaling is observed, and goes up to electron scales. However, in our simulation the \textit{exp} model works even if we start the fit directly in correspondence of the spectral break at ion scales. A possible explanation for this difference with respect to SW observations is that the variability in the slope of the SW magnetic spectrum around the ion scale spectral break seems to be induced by instabilities that inject energy around $k\,d_i\!\simeq\!1$ \citep{Alexandrova2013}. Such instabilities are triggered by temperature anisotropies generated by the SW expansion \citep{bale2009magnetic,Alexandrova2013}, an effect that is not present in our simulation and may thus explain the aforementioned discrepancy. Another factor influencing the shape of the magnetic spectrum at ion scales is the ion plasma beta $\beta_i$. In the study of \citet{Franci2016} conducted using hybrid simulations, it is shown that the magnetic field spectrum at sub-ion scales becomes shallower as $\beta_i$ increases, reaching the $\sim\!k^{-2.8}$ power law already around $k\,d_i\!\simeq\!1$ (where the spectral break is observed) when $\beta_i\!>\!1$. This represents another possible reason why in our case the \textit{exp} models works already from the spectral break downwards and with a scaling exponent close to $-2.8$. The reduced mass ratio $m_i/m_e\!=\!100$ we used is also another parameter influencing the properties of the spectra. Using a reduced mass ratio implies that the range between ion and electron scales (where the $k^{-2.8}$ scaling is expected) becomes narrower (but still reasonably separated) and thus electron kinetic effects become relevant at scales larger than in a real plasma. Therefore, to finally confirm our finding, simulations with a more realistic mass ratio are needed in order to properly separate ion and electron scales. However, such simulations have a computational cost that makes them very challenging nowadays, so the mass ratio we used is at the moment a reasonable compromise and we speculate that if the \textit{exp} models is universal, a realistic mass ratio would imply a shift of the exponential range towards larger wavenumbers (since a larger mass ratio implies a smaller $\rho_e$). Another important point to discuss is the comparison of our results with observational studies supporting the idea of the double power law, in contrast to the \textit{exp} model. In particular, in \citet{chen2017nature} the magnetic field spectrum measured in the magnetosheath is shown to be consistent with a double power law around electron scales. The authors propose a theoretical model based on inertial kinetic Alfvén waves to explain the steepening observed for $k\,d_e\!\geqslant\!1$. This model assumes that $\beta_e\!\ll\!\beta_i\!\simeq\!1$ (from which it follows that effects related to the electron pressure are weak) and predicts a scale for the spectral break corresponding to $k^2\,\rho_e^2\!\sim\!\beta_i(2+\beta_i)(T_e/T_i)^2$ (where $T_e$ and $T_i$ are the electron and ion temperatures). In our simulation these conditions on the ion and electron plasma betas are not satisfied since both $\beta_i$ and $\beta_e$ are larger than $1$ and the electron pressure plays a key role in the dynamics of the system at sub-ion scales (as discussed in the analysis of the filtered energy conversion channels). Moreover, the model would predict a spectral break at scales $k\,\rho_e\!\sim\!2.85$ in our simulation (at $t\!=\!650\,\Omega_e^{-1}$), corresponding to $k\,d_i\!\sim\!17.4$, laying inside the exponential range. Therefore, the model proposed in \citet{chen2017nature} is not suited to describe our results. Nonetheless, this comparison shows that the features of the magnetic field spectrum around electron scales may depend on the balance between inertial and kinetic effects of the electrons, i.e. on the value of $\beta_e$. We speculate that for low $\beta_e$ (as in the observations reported by \citet{chen2017nature}) electron inertial effects may cause the development of a power law around $k\,d_e\!\simeq\!1$ while for high $\beta_e$ (as in our simulation) electron kinetic effects may generate an exponential decay around $k\,\rho_e\!\simeq\!1$. However, this point cannot be properly assessed with a single simulation and a parametric study with respect to $\beta_e$ is needed.

Furthermore, we found that the \textit{exp} model represents a good description also for the electron velocity spectrum at sub-ion scales, with a scaling exponent $\alpha_{u_e}\!\simeq\!0.94$ and a characteristic length $\lambda_{u_e}\!\simeq\!0.87\,\rho_e$, very close to the scale $\lambda_B\!\simeq\!\rho_e$ found for the magnetic field spectrum. On the other hand, no exponential range was observed in the ion velocity spectrum that does not extend much into electron scales and drops as a steep power law $\sim k^{-3.25}$ at sub-ion scales. This scaling is remarkably consistent with SW observations reported in \citet{vsafrankova2016power} where an average spectral index between $-3.2$ and $-3.3$ is found for the ion velocity spectrum when $3\!\lesssim\!\beta_i\!\lesssim\!16$ (as in our case where $\beta_i\!=\!8$ at the beginning of the simulation and grows up to $\beta_i\!\simeq\!14$ at $t\!=\!650\,\Omega_e^{-1}$, due to the ion heating).

We investigated the dynamics responsible for the development of the turbulence from large to sub-ion scales by analyzing the filtered ion and electron energy conversion channels. Our analysis outlines the major role played by the electrons with respect to the ions in driving the magnetic field dynamics at sub-ion scales. We have shown that in our simulation there are two possible channels accounting for the transfer of e.m. energy from large to sub-ion scales: a direct scale-to-scale transfer described by the cross scale flux of e.m. energy $\Pi_i^{bb}\!+\!\Pi_e^{bb}$, and an indirect transfer lead by the electrons that first subtract energy from the e.m. field at large scales (converting it into electron fluid flow energy), then transfer it to sub-ion scales and finally give it back to the e.m. field (see Fig.~\ref{EnergyBalance}). The latter electron-driven mechanism is way more efficient than the direct scale-to-scale transfer in channelling the e.m. energy from large to sub-ion scales. On the other hand, the ions do not contribute to the transport of e.m. energy to small scales and we observe only a weak conversion of e.m. energy into ion fluid flow energy at sub-ion scales, which possibly explains why the ion velocity spectrum quickly drops at kinetic scales (consistently with the fact that ions are expected to decouple from the magnetic field at those scales). Therefore, the e.m. field dynamics at kinetic scales is mainly supported by the electrons and this may explain the similarities in the magnetic field and electron velocity spectra (in particular, the presence of an electron scale exponential range that is instead absent in the ion velocity spectrum). We also observed that the electron-driven transfer of e.m. energy acts since the beginning of the simulations and stops at fully developed turbulence, when the electron heating becomes more efficient than the cross-scale flux of fluid flow energy coming from large scales. At that point also the electrons start to take energy from the e.m. field and eventually convert it into internal energy. The pressure-strain decomposition shows that this sub-ion scale electron heating is basically incompressible since it is almost entirely determined by the Pi-D interaction (see Fig.~\ref{PS_e_>}). This transition to a dissipative regime at fully developed turbulence is most likely caused by the lack of a large scale forcing that maintains the turbulent cascade (i.e. we are dealing with freely decaying turbulence). This implies that there is no large scale source replacing the energy that is being dissipated at small scales, so once most of the large scale energy has reached kinetic scales, the transfer of energy to sub-ion scales becomes less efficient than dissipation and we observe the aforementioned transition to a dissipative dynamics. Another point that needs to be discussed is the role of the ions at large scales. In Fig.~\ref{ECC_i_<} we saw that the ions overall lose energy at $k\,d_i\!<\!1.5$ throughout the simulation, both via the pressure-strain interaction and the e.m. work. This behavior could depend on the fact that with our initialization most of the initial energy is contained in the ion fluid flow energy that is larger than both the e.m. energy and the electron fluid flow energy (as seen in Fig.~\ref{Energy}). As a consequence, the system may tend to reduce this energy gap by converting the ion fluid flow energy into e.m. energy and ion internal energy. Therefore, we do not claim that the global energy balance observed in our simulation is general and universal because it may depend on various factors such as the way the turbulence is initialized at large scales and the strength of dissipation, determined by the plasma beta (a higher plasma beta implies a stronger pressure-strain interaction and thus more efficient dissipation, as discussed also in \citet{parashar2018dependence}). However, a parametric study with respect to the level of initial fluctuations and the ion and electron plasma betas is beyond the scope of the present paper and will be discussed in future works. Nonetheless, the analysis performed on our simulation shows that the filtered energy conversion channels are indeed a very useful and powerful tool to describe and study the turbulent energy cascade since they allow to accurately track the path that the energy follows in its way from large to small scales. Finally, another limitation of our work that is worth mentioning is represented by the limited box size and the 2D geometry of our simulation. Additional studies with larger and possibly 3D simulation domains will be needed in order to have results that are finally comparable with satellite observation.

\begin{acknowledgements}
      This work has received funding from the KULeuven Bijzonder Onderzoeksfonds (BOF) under the C1 project TRACESpace, from the European Union’s Horizon 2020 research and innovation program under grant agreement No. 776262 (AIDA). Computing has been provided by the Flemish Supercomputing Center (VSC) and by the PRACE Tier-0 program. Numerical simulations have been performed on SuperMUC-NG, hosted by the Leibniz Supercomputing Centre (Germany), under the PRACE project.\\
      The simulation data-set (KUL\_MSH) used in this work is available at Cineca on the AIDA-DB database. In order to access the meta-information and the link to the raw simulation data see the tutorial at http://aida-space.eu/AIDAdb-iRODS.
\end{acknowledgements}

\bibliography{peppe}
\bibliographystyle{aa}

%
%

\end{document}